\documentclass[fontsize=12pt,paper=a4]{article}
\usepackage[english]{babel}
\usepackage{amsmath}
\usepackage{amssymb}
\usepackage{graphicx}
\usepackage{hyperref}

\newcommand{\hc}[1]{#1^{\dagger}}
\newcommand{\T}[1]{#1^{T}}
\newcommand{\abs}[1]{|#1|}
\newcommand{\sgn}{\operatorname{sgn}}
\newcommand{\tr}{\operatorname{tr}}

\newcommand{\adj}{\operatorname{adj}}

\author{Kristjan Kannike\\
{\footnotesize National Institute of Chemical Physics and Biophysics, R\"avala 10, 10143 Tallinn, Estonia}
\\
{\footnotesize kannike@cern.ch}
}

\date{} 

\title{Vacuum Stability of a General Scalar Potential of a Few Fields}

\begin{document}
\maketitle

\begin{abstract}
We calculate analytical vacuum stability or bounded from below conditions for general scalar potentials of a few fields. After a brief review of copositivity, we  show how to find positivity conditions for more complicated potentials. We discuss the vacuum stability conditions of the general potential of two real scalars, without and with the Higgs boson included in the potential. As further examples, we give explicit vacuum stability conditions for the two Higgs doublet model with no explicit CP breaking, and for the $\mathbb{Z}_{3}$ scalar dark matter with an inert doublet and a complex singlet. We give a short overview of positivity conditions for tensors of quartic couplings via tensor eigenvalues. A Mathematica notebook is included with the source files.
\end{abstract}

\tableofcontents

\newpage

\section{Introduction}

A scalar potential has to be bounded from below to make physical sense. In the Standard Model (SM), it simply means that the self-coupling of the Higgs boson has to be positive. In an extended model with more scalar fields, the potential has to be bounded from below -- the vacuum has to be stable -- in the limit of large field values in all possible directions of the field space. In this limit, any terms with dimensionful couplings -- mass or cubic terms -- can be neglected in comparison with the quartic part of the scalar potential.%
\footnote{The requirement of strong stability means demanding that the quartic part of the potential $V_{4} > 0$ as the fields $\varphi_{i} \to \infty$, whereas $V_{4} \geqslant 0$ gives stability in the marginal sense and there can be flat directions (then the quadratic or mass squared part of the potential \emph{has} to be positive and there cannot be any cubic terms). For simplicity, we give conditions for strong stability, which in practice means making inequalities strict.}

In quantum field theories, scalar couplings change with energy due to the renormalisation group running. The vacuum stability conditions may be satisfied at some scales and not satisfied at others. Checking the vacuum stability of the tree level potential with running couplings can help to determine the scale of validity of a given model. On the other hand, in models with classical scale invariance \cite{Bardeen:1995kv}, where tree-level mass terms are absent, violation of vacuum stability conditions at a \emph{finite} field range can be used to produce minima and induce symmetry breaking via dimensional transmutation as in the Coleman-Weinberg mechanism \cite{Coleman:1973jx}. 

The remarkable Tarski-Seidenberg theorem implies that the question of whether the vacuum is stable or not for given values of the scalar couplings is in principle always decidable. Nevertheless, the general problem of finding whether a given polynomial is non-negative is an NP-hard problem if the degree of the polynomial is at least four \cite{springerlink:10.1007/BF02592948}, which is the case for renormalisable scalar potentials in four dimensions.

The most general quartic potential of real scalars is, of course,
\begin{equation}
  V(\phi) = \lambda_{ijkl} \phi_i \phi_j \phi_k \phi_{l},
\label{eq:V:general}
\end{equation}
where the coupling tensor $\lambda_{ijkl}$ can be always made completely symmetric under any exchange of the indices. Alas, relatively simple complete conditions of the positivity of the potential \eqref{eq:V:general} can be given only in the case of two fields (Sect.~\ref{sec:two:real:scalars}), or of three fields, if the potential is biquadratic in one of them (Sect.~\ref{sec:two:real:scalars:and:Higgs}).

Since $V(\phi)$ is a homogenous quartic polynomial, scaling the fields by a positive real constant $c$ gives $V(c \, \phi) = c^{4} V(\phi)$ and does not affect vacuum stability.\footnote{We can even scale each field $\phi_{i}$ separately by a different positive coefficient. In particular, we can scale it by $\phi_{i} \to \phi_{i}/\lambda_{i}^{\frac{1}{4}}$, where $\lambda_{i}$ is its self-coupling, and make the coefficient of $\phi_{i}^{4}$ equal to unity for the purpose of calculating positivity conditions (this only holds at tree level or at a fixed energy scale).} Therefore, we can write the quartic potential as
\begin{equation}
   V(\phi) = V(\hat{\phi}) \, r^{4},
\end{equation}
where $\hat{\phi}^{2} = 1$ and $r \geqslant 0$. We see that if $V(\hat{\phi})$ is negative for some $\hat{\phi}$, then the potential tends to negative infinity as $r \to \infty$ and the vacuum is not stable.

Thus, to determine whether a potential is bounded from below in the limit of large field values, we can minimise its quartic part on a unit hypersphere,
enforced by a Lagrangian multiplier $\lambda$:
\begin{equation}
    V(\phi, \lambda) = V(\phi) + \frac{\lambda}{2} \left( 1 - \phi^{2} \right),
\label{eq:V:Lagrange:multiplier}
\end{equation}
which yields the stationary point equations
\begin{equation}
  \frac{\partial V(\phi)}{\partial \phi_{i}} = \lambda \phi_{i}, \quad \phi^{2} = 1.
  \label{eq:eqs:Lagrange:multiplier}
\end{equation}
Notice that $\lambda = 4 V_{\text{min}}$ since if we write the constraint on $\phi$ as $g(\phi, c) = c^{2} - \phi^{2}$, then on one hand $d V_{\text{min}}/d c = \lambda c$, while on the other hand $d V_{\text{min}}/ d c = 4 c^{3} V_{\text{min}}$ with $c = 1$ in the end.

Global or gauge symmetries of the potential may help to simplify the problem. An important special case is given by quartic potentials that are biquadratic in fields and have the form
\begin{equation}
  V = \lambda_{ij} \phi_i^{2} \phi_j^{2}.
\label{eq:V:biquadratic}
\end{equation}
The couplings $\lambda_{ij}$ can be written as a matrix on the basis of $\phi_{i}^{2}$. Since the squares of real fields are non-negative, the natural domain of such potentials is not $\mathbb{R}^{n}$ but the non-negative orthant $\mathbb{R}_{+}^{n}$. Positivity on $\mathbb{R}_{+}^{n}$ or copositivity (short for `conditional positivity') was introduced in \cite{Motzkin:1952aa}.\footnote{See \cite{Hiriart-Urruty:2010fk} for a good review of copositive matrices.} Copositivity has found wide use in the field of convex optimisation, and was first used to derive vacuum stability conditions in \cite{Kannike:2012pe}.  The set of copositive matrices is larger than  and includes the familiar set of positive semidefinite matrices.\footnote{For a recent review on positive semidefinite matrices, see \cite{Hiriart-Urruty:2012qy}.} While the positive definite part of the parameter space can be easily found via Sylvester's criterion \cite{Gilber:1991:PDM:115430.115439}, the criteria for copositivity are more involved, but definite analytic procedures exist to compute them. 

Even if the fields are gauge multiplets, any potential can be written in terms of field magnitudes and orbit space variables \cite{Frautschi:1981jh,Kim:1981jj,Kim:1981xu,Kim:1982dn,Kim:1983mc}. In many cases the potential is a monotonous function of orbit space parameters and its minimum occurs on the boundary of the orbit space. The symmetries of the potential may restrict the variables to a more complicated space such as the `future light cone' orbit space of the two Higgs doublet model (2HDM) \cite{Sartori:2003ya,Maniatis:2006fs,Ivanov:2006yq,Ivanov:2015nea} or the similar orbit spaces of the 3HDM \cite{Maniatis:2014oza} and NHDM \cite{Maniatis:2015gma}. The importance of taking the orbit space into account properly can be seen, for example, in the case of the type II seesaw: the vacuum stability conditions calculated in \cite{Arhrib:2011uy} were somewhat too strong, because the orbit space parameters do not vary independently \cite{Bonilla:2015eha}. 

In general, the conditions for a potential to be bounded from below can be expressed in many ways. 
It may be possible to produce conditions for vacuum stability that are analytical but of considerable length. For more complicated potentials with several fields, one has to resort to numerics, for which the methods we present can still be useful for reducing the parameter space to scan over. The purpose of the paper is to introduce into the `toolbox' of calculating vacuum stability conditions some methods that are specific, but useable in many practical cases, and others that are more complicated but also more general. While recent mathematical literature is concerned with approximate methods of finding positivity for polynomials of many variables, particle physics models usually deal with a few scalar fields and analytical solutions may afford more insight.

On numerous occasions, the new addition to the scalar sector consists of just a couple of real scalar singlets, often in the guise of a complex singlet. In this case the vacuum stability reduces to the problem of positivity of a general quartic polynomial. For comparison, we also derive the vacuum stability conditions in another form, using the Sylvester criteria for the matrix of scalar couplings. And, of course, no low energy scalar potential is complete without the Standard Model Higgs doublet, which we learn to include as well. Similar conditions can be derived e.g. for the 2HDM, where the potential can be considered to be a quartic polynomial in magnitudes of fields, or for more complicated models, such as the $\mathbb{Z}_{3}$ scalar dark matter \cite{Belanger:2012vp,Belanger:2014bga} with an inert doublet and a complex singlet. To our knowledge, the results for the potentials of the two singlets (and the Higgs) and for the $\mathbb{Z}_{3}$ scalar dark matter are new. As for the 2HDM with real couplings, our results are in shorter form than similar results in the literature \cite{Eriksson:2009ws}. 

In addition, we reconsider the notion of copositivity of matrices and discuss its relation to orbit space variables. In more complicated situations, tests of (co)positi\-vity in terms of eigenvalues of the \emph{tensor} of scalar couplings can help. Similarly to a positive matrix, a tensor is positive-definite if its eigenvalues associated with its real eigenvectors are positive.

In Sect.~\ref{sec:orbit:spaces} we give a brief review of orbit spaces and copositivity. In Sect.~\ref{sec:two:real:scalars} we give the conditions for the general potential of two real scalars to be positive. It is not too hard to include the SM Higgs doublet into the potential in Sect.~\ref{sec:two:real:scalars:and:Higgs}. In Sect.~\ref{sec:2hdm} we derive vacuum stability conditions for the 2HDM with no explixit CP-breaking. Sect.~\ref{sec:z3:scalar:DM} provides another illustration in the vacuum stability conditions for $\mathbb{Z}_{3}$ scalar dark matter. In Sect.~\ref{sec:tensor:eigenvalues} we introduce tensor eigenvalues as a way to determine the vacuum stability conditions for a most general scalar potential. We conclude in Sect.~\ref{sec:conclusions}. A Mathematica notebook with the conditions for all examples is included with the \LaTeX\ source of the paper.

\section{ Copositivity \emph{\&} Orbit Spaces}
\label{sec:orbit:spaces}

A scalar potential \eqref{eq:V:biquadratic} biquadratic in fields is bounded from below if the matrix of couplings $\lambda_{ij}$ is copositive \cite{Kannike:2012pe}. Even if the fields are higher multiplets under a gauge group, any potential can be written in terms of squares of field magnitudes and a few dimensionless orbit space variables. 

\subsection{Copositivity}
\label{sec:cop}

The criteria to determine whether a matrix is positive in the usual sense are well established. \emph{A symmetric matrix $A$ is said to be \emph{positive semidefinite} if the quadratic form $\T{x} A x \geqslant 0$ for all vectors $x$ in $\mathbb{R}^{n}$. A symmetric matrix $A$ is \emph{positive definite} if the inequality is strict, $\T{x} A x > 0$ for any non-zero vector $x$ in $\mathbb{R}^{n}$.} A matrix $A$ is positive (semi)definite if and only if (i) the eigenvalues of $A$ are positive (non-negative), (ii) the principal minors of $A$ are positive (non-negative), or (iii) the principal invariants of $A$ are positive (non-negative). The principal minors of $A$ are determinants of the principal submatrices, which are obtained by deleting $k$ rows and columns from $A$ in a symmetric way, i.e. both the $i_{1}, \ldots, i_{k}$ rows and the $i_{1}, \ldots, i_{k}$ columns are deleted. The largest principal submatrix of $A$ is $A$ itself.

On the other hand, copositive matrices are demanded to be positive not for all vectors in the $\mathbb{R}_{n}$, but only on positive vectors in $\mathbb{R}_+^n$. \emph{A symmetric matrix $A$ is \emph{copositive} if the quadratic form $\T{x} A x \geqslant 0$ for all vectors $x \geqslant 0$ in the non-negative orthant $\mathbb{R}_+^n$.} (The notation $x \geqslant 0$ means that $x_i \geqslant 0$ for each $i = 0, \ldots, n$.)  \emph{A symmetric matrix $A$ is \emph{strictly copositive} if the quadratic form $\T{x} A x > 0$ for all vectors $x > 0$ in the non-negative orthant $\mathbb{R}_+^n$.}

For matrices of low order the copositivity conditions are relatively simple.
A symmetric matrix $A$ of order 2 is copositive if and only if \cite{KP198379}
\begin{equation}
  a_{11} \geqslant 0, \quad a_{22} \geqslant 0, \quad  a_{12} + \sqrt{ a_{11} a_{22} } \geqslant 0. \label{eq:A:2:copos:nondiag:12}
\end{equation}
A symmetric matrix $A$ of order 3 is copositive if and only if \cite{Hadeler198379,Chang1994113}
\begin{align}
  a_{11} &\geqslant 0, & a_{22} &\geqslant 0, & a_{33} &\geqslant 0,
  \notag
  \\
  \bar{a}_{12} = a_{12} + \sqrt{ a_{11} a_{22} } &\geqslant 0,
  &
  \bar{a}_{13} = a_{13} + \sqrt{ a_{11} a_{33} } &\geqslant 0,
  &
  \bar{a}_{23} = a_{23} + \sqrt{ a_{22} a_{33} } &\geqslant 0,
  \label{eq:A:3:copos:crit}
  \\
  \span \span \span \span \sqrt{a_{11} a_{22} a_{33}} + a_{12} \sqrt{a_{33}} + a_{13} \sqrt{a_{22}} + a_{23} \sqrt{a_{11}} 
  + \sqrt{2 \bar{a}_{12} \bar{a}_{13} \bar{a}_{23}} &\geqslant 0.
  \notag
\end{align}

The Cottle-Habetler-Lemke theorem \cite{Cottle1970295} provides a practical way to find analytical copositivity conditions for matrices of low order. \emph{Let the order $n-1$ submatrices of a real symmetric matrix $A$ of order $n$ be copositive. Then $A$ is \emph{not} copositive if and only if
\begin{equation}
  \det A < 0 \quad \land \quad \adj A \geqslant 0,
\end{equation}
or $A$ is copositive if and only if
\begin{equation}
  \det A \geqslant 0 \quad \lor \quad (\adj A)_{ij} < 0 \text{ for some } i,j.
\end{equation}
}
The adjugate of $A$ is the transpose of the cofactor matrix of $A$:
\begin{equation}
  (\adj A)_{ij} = (-1)^{i+j} M_{ji},
\end{equation}
where $M_{ij}$ is the $(i,j)$ minor of $A$, the determinant of the submatrix resulting from deleting the $i$th row and $j$th column of $A$.

Another general way to test copositivity is Kaplan's test \cite{Kaplan:2000:TCM}: \emph{A symmetric matrix $A$ is copositive if and only if every principal submatrix of $A$ has no eigenvector $v \geqslant 0$ with associated eigenvalue $\lambda \leqslant 0$.}

We see that while the positivity of the matrix can be checked via its eigenvalues and the direction of its eigenvectors is irrelevant in this case, to check for copositivity at worst all the eigenvalues and eigenvectors of all principal submatrices of the matrix have to be calculated.

Copositivity with respect to a closed cone $K$ or $K$-copositivity means
\begin{equation}
  \T{x} A x \geqslant 0 \quad \forall ~ x \in K.
\label{eq:cone:cop}
\end{equation}

If the cone $K$ is polyhedral, it can be represented in the form $K = \{G x \, | \, x \in \mathbb{R}_{+}^{p} \}$, where $G$ is a real matrix whose columns $\{ u_{1}, \ldots, u_{n} \}$ are positively linearly independent vectors in $\mathbb{R}^{n}$ that map the extremal rays of $\mathbb{R}_{+}^{n}$  (basis vectors) into the extremal rays of the cone $K$. In such a case, the condition \eqref{eq:cone:cop} takes the form \cite{Hiriart-Urruty:2010fk}
\begin{equation}
  \T{x} \T{G} A G x \geqslant 0 \quad \forall ~ x \in R_{+}^{p}.
\label{eq:cone:cop:via:cop}
\end{equation}
Therefore to check whether $A$ is copositive on $K$, we can check the usual copositivity of $\T{G} A G$ on $\mathbb{R}_{+}^{n}$.

Kaplan's test can also be generalised to copositivity on a closed cone $K$ by requiring that every principal submatrix of $A$ have no eigenvector $v \in K$ with associated eigenvalue $\lambda \leqslant 0$.

\subsection{Orbit Spaces}
\label{sec:orbit}

Because the scalar potential is a homogenous polynomial of fields, the question whether the potential is bounded from below does not depend on the norm of the field for a single scalar field (e.g. the Higgs in the SM) and depends only on ratios of the norms for multiple fields. Orbit spaces for different gauge groups and potentials were actively studied in the 1980s, especially in the context of Grand Unified Theories. We give a short review with references to some works. The space of magnitudes of fields is the non-negative orthant $\mathbb{R}_{+}^{n}$ and hence the problem of positivity of the potential can be put in terms of copositivity of the matrix (or tensor) of quartic couplings on the basis of norms of fields, with a few orbit space parameters to minimise over.
 
Orbit spaces in the context of spontaneous symmetry breaking are described in \cite{Abud:1983id,Abud:1981tf}. The method of orbit spaces to minimise potentials is detailed in \cite{Kim:1981xu}. The case of the adjoint representation of $SU(N)$, in particular of the $\mathbf{24}$ of $SU(5)$, is detailed in \cite{Kim:1981jj}. An illustration for a potential of two scalars in different representations is given in \cite{Frautschi:1981jh} for the orbit space for the quartic potential of the $\mathbf{5}$ and the $\mathbf{24}$ representation of $SU(5)$. In \cite{Kim:1982dn} the case of $SO(N)$ with adjoint + vector representation is analysed. For several groups, the orbit spaces for simple potentials are described in \cite{Kim:1983mc}. 

We present a short outline of the method for a potential of one scalar multiplet. In our cursory review we will heed to the presentation of Jai Sam Kim \cite{Kim:1981xu}. For a theory with the non-Abelian gauge group $G$, the quartic potential for a scalar $\phi$ in an $n$-dimensional irreducible representation $\mathbf{R}$ of $G$ can be written as
\begin{equation}
  V(\phi) = \lambda_{\phi} (\phi_{i}^{*} \phi_{i})^{2} 
  + \lambda'_{\phi} f_{ijkl} \phi_{i}^{*} \phi_{j} \phi_{k}^{*} \phi_{l}
  + \lambda''_{\phi} g_{ijkl} \phi_{i}^{*} \phi_{j} \phi_{k}^{*} \phi_{l} + \ldots,
\label{eq:V:higher:phi}
\end{equation}
where $f$ and $g$ specify different gauge invariant contractions of indices.
The potential $V(\phi)$ is invariant under a group transformation
\begin{equation}
  \phi_{j} = T(\theta)_{ji} \phi_{i},
\end{equation}
where $T(\theta)$ is an $n$-dimensional matrix representing a group element. In general
\begin{equation}
  T(\theta) = e^{- i \theta_{L} X_{L}},
\end{equation}
where $X_{L}$ are the generators of the group $G$ and $\theta_{L}$ are parameters that specify the group element.

Because in general the multiplet $\phi$ has many components, it is hard to solve the minimisation equations \eqref{eq:eqs:Lagrange:multiplier} for them. Moreover, there is a degeneracy of  the components of $\phi$ that give the same minimum of $V$. A gauge transformation rotates the components of $\phi$ while leaving the value of the potential unchanged.

More formally, the orbit of a particular $\phi$ with constant components (such as a vacuum expectation value) is the set of states $\phi_{\theta} = T(\theta) \phi$ with $T(\theta)$ an element of $G$. It can be shown that all the states $\phi_{\theta}$ respect the same group, the little group of the orbit, as $\phi$ does. If the group is unitary, then all the states $\phi_{\theta}$ have the same norm $\phi^{\dagger} \phi$. The  set of orbits that respect the same little group is called the stratum of the little group. Therefore we have to look for the orbit --  and its little group -- that minimises the potential. 

Orbits of $\phi$ are specified by invariant polynomials $P(\phi)$ \cite{Weyl,Dieudonne19701,Luna1976}. There is a basis set of invariant polynomials $I_{a}(\phi)$ such that every invariant polynomial $P(\phi)$ can be expressed as a polynomial in the polynomial basis: $P(\phi) = \bar{P}[I_{a}(\phi)]$. Each representation $\mathbf{R}$ has  a different number $\ell$ of basic invariants. An orbit can be pictured as a point in the $\ell$-dimensional space of $I_{a}$.

The magnitude of $\phi$ is irrelevant to minimising the potential to find the vacuum stability conditions. The strata can be specified by dimensionless ratios of invariants, e.g.
\begin{equation}
  \alpha_{1} = \frac{f_{ijkl} \phi_{i}^{*} \phi_{j} \phi_{k}^{*} \phi_{l}}{(\phi_{i}^{*} \phi_{i})^{2}}.
\end{equation}
These dimensionless ratios are called orbit parameters and can be thought of as a set of angles. The  potential \eqref{eq:V:higher:phi} can then be written as
\begin{equation}
   V(\phi) = \left[ \lambda_{\phi} + \lambda'_{\phi} \alpha_{1}(\hat{\phi}) + \lambda''_{\phi} \alpha_{2}(\hat{\phi}) + \ldots \right] \abs{\phi}^{4},
\end{equation}
where
\begin{equation}
  \abs{\phi}^{2} = \phi_{i}^{*} \phi_{i}, \quad \hat{\phi}_{i} = \frac{\phi_{i}}{\abs{\phi}}.
\end{equation}
The potential $V(\phi)$ is bounded from below if
\begin{equation}
  \lambda_{\phi} + \lambda'_{\phi} \alpha_{1}(\hat{\phi}) + \lambda''_{\phi} \alpha_{2}(\hat{\phi})  
  + \ldots > 0 \quad
  \text{for any $\alpha_{i}(\hat{\phi})$}.
\label{eq:min:cond:one:irrep}
\end{equation}
Therefore, we have to minimise the potential with respect to the orbit space parameters. It is obvious that for any $\hat{\phi}$, the range of $\alpha_{i}$ is bounded from below and above: $\alpha_{i\text{min}} \leqslant \alpha_{i} \leqslant \alpha_{i\text{max}}$. We have to calculate the orbit space -- the physical region in the orbit space parameters $\alpha_{i}$. Because $\lambda_{\phi} + \lambda'_{\phi} \alpha_{1} + \lambda''_{\phi} \alpha_{2} = C$ describes a line in the orbit space, the minimum of the potential is on the boundary of the orbit space, in particular it can be on a cusp of the boundary curve.

Similar considerations apply for more than two orbit space parameters and for several scalars in different representations. For many more details, we refer the interested reader to the works cited. 

\section{Vacuum Stability of the Scalar Potential of Two Real Scalars}
\label{sec:two:real:scalars}

\subsection{Vacuum Stability Conditions from Positivity of a Quartic Polynomial}

The most general scalar potential of two real scalar fields $\phi_{1}$ and $\phi_{2}$ is\footnote{Of course, the most general potential of the real and imaginary components of a complex singlet $S = \phi_{1} + i \phi_{2}$ can be written in the same form. If $\lambda_{31} = \lambda_{13} = 0$, the potential has a CP symmetry. Therefore $\lambda_{31}$ and $\lambda_{13}$ could be naturally small.} 
\begin{equation}
  V(\phi_{1}, \phi_{2}) = \lambda_{ij} \phi_{1}^{i} \phi_{2}^{j} 
  = \lambda_{40} \phi_{1}^{4} + \lambda_{31} \phi_{1}^{3} \phi_{2} + \lambda_{22} \phi_{1}^{2} \phi_{2}^{2} + \lambda_{13} \phi_{1} \phi_{2}^{3} + \lambda_{04} \phi_{2}^{4}.
\label{eq:V:general:2:fields}
\end{equation}
Here we use the same notation $\lambda_{ij}$ in a different way than in eq. \eqref{eq:V:biquadratic} and copositivity, but the meaning should be clear from the context.

Set either $\phi_{1}$ or $\phi_{2}$ to zero, and it follows that the self-couplings $\lambda_{40}$ and $\lambda_{04}$ have to be positive in order for $V$ to be positive.%
\footnote{Note that we can scale the fields by $\phi_{1} \to \phi_{1}/\lambda_{40}^{1/4}$ and $\phi_{2} \to \phi_{2}/\lambda_{04}^{1/4}$, so in effect there are three independent parameters to consider.}
When both fields are non-zero, we can divide the potential $V$ by $\phi_{2}^{4}$ and choose the ratio $x = \phi_{1}/\phi_{2}$ as the new variable (equivalently, set $\phi_{2} = 1$ to dehomogenise $V$), reducing the question of vacuum stability of the potential to that of the positivity of a general quartic polynomial of one variable,
\begin{equation}
  P = a_{4} x^{4} + a_{3} x^{3} + a_{2} x^{2} + a_{1} x + a_{0},
\label{eq:general:quartic}
\end{equation}
which obviously is positive if it has no real roots and $a_{4} > 0$ and $a_{0} > 0$.

The nature of the roots of $P$ can be determined by considering its discriminant
\begin{equation}
\begin{split}
  D &= 256 a_{0}^{3} a_{4}^{3} - 4 a_{1}^{3} a_{3}^{3} - 27 a_{0}^{2} a_{3}^{4} 
  + 16 a_{0} a_{2}^{4} a_{4} - 6 a_{0} a_{1}^{2} a_{3}^{2} a_{4} - 27 a_{1}^{4} a_{4}^{2} \\
  & - 192 a_{0}^{2} a_{1} a_{3} a_{4}^{2} - 4 a_{2}^{3} (a_{0} a_{3}^{2} + a_{1}^{2} a_{4}) 
  + 18 a_{2} (a_{1} a_{3} + 8 a_{0} a_{4}) (a_{0} a_{3}^{2} + a_{1}^{2} a_{4}) \\
  &+ a_{2}^{2} (a_{1}^{2} a_{3}^{2} - 80 a_{0} a_{1} a_{3} a_{4} - 128 a_{0}^{2} a_{4}^{2}),
\end{split}
\end{equation}
and two additional polynomials of its coefficients,
\begin{equation}
  Q = 8 a_{2} a_{4} - 3 a_{3}^{2}, 
  \quad
  R = 64 a_{0} a_{4}^{3} + 16 a_{2} a_{3}^{2} a_{4} - 16 a_{4}^{2} (a_{2}^{2} + a_{1} a_{3}) 
  - 3 a_{3}^{4}.
\end{equation}
The condition for $P$ to have only complex roots is \cite{Rees:1922,Lazard:1988}
\begin{equation}
  D > 0 \land (Q > 0 \lor R > 0).
  \label{eq:one:variable:polynomial:positivity}
\end{equation}
In the marginal case $D = 0$ (which we can usually ignore), the conditions 
\begin{equation}
Q > 0, \quad R = 0, \quad S = a_{3}^{3} + 8 a_{1} a_{4}^{2} - 4 a_{4} a_{3} a_{2} = 0
  \label{eq:one:variable:polynomial:positivity:marginal}
\end{equation} must hold.

For $V(\phi_{1}, \phi_{2})/\phi_{2}^{4}$, the discriminant $D$ and the polynomials $Q$ and $R$ are  given by
\begin{align}
  D &= 
  256 \lambda_{40}^{3} \lambda_{04}^{3} - 4 \lambda_{31}^{3} \lambda_{13}^{3}
  - 27 \lambda_{31}^{4} \lambda_{04}^{2} + 16 \lambda_{40} \lambda_{22}^{4} \lambda_{04}
  - 6 \lambda_{40} \lambda_{31}^{2} \lambda_{04} \lambda_{13}^{2}  
  - 27 \lambda_{40}^{2} \lambda_{13}^{4} \notag \\
  &  - 192 \lambda_{40}^{2} \lambda_{31} \lambda_{04}^{2} \lambda_{13} 
  - 4 \lambda_{22}^{3} (\lambda_{31}^{2}  \lambda_{04} + \lambda_{40} \lambda_{13}^{2}) 
  + 18 \lambda_{22} (8 \lambda_{40} \lambda_{04} + \lambda_{31} \lambda_{13}) 
  \label{2:real:2:field:discriminant}
  \\
  &\times (\lambda_{31}^{2} \lambda_{04} + \lambda_{40} \lambda_{13}^{2}) 
  + \lambda_{22}^{2} (\lambda_{31}^{2} \lambda_{13}^{2} - 80 \lambda_{40} \lambda_{31} \lambda_{04} \lambda_{13} - 128 \lambda_{40}^{2} \lambda_{04}^{2}), \notag
  \\
  Q &= 8 \lambda_{40} \lambda_{22} - 3 \lambda_{31}^{2}, 
  \label{2:real:2:field:P}
  \\
  R &= 64 \lambda_{40}^{3} \lambda_{04} + 16 \lambda_{40} \lambda_{22} \lambda_{31}^{2} 
  - 16 \lambda_{40}^{2} (\lambda_{22}^{2} + \lambda_{31} \lambda_{13}) - 3 \lambda_{31}^{4},
  \label{2:real:2:field:D:Q:R}
\end{align}
and the vacuum stability conditions for $V(\phi_{1}, \phi_{2})$ are given by%
\begin{equation}
  \lambda_{40} > 0, \quad \lambda_{04} > 0, \quad D > 0 \land (Q > 0 \lor R > 0).
  \label{eq:2:real:2:field:vac:stab}
\end{equation}
For the record, the polynomial $S = \lambda_{13}^{3} + 4 \lambda_{40} (2 \lambda_{40} \lambda_{31} - \lambda_{22} \lambda_{13})$.

Of course, we could as well divide by $\phi_{1}^{4}$ and choose $\phi_{2}/\phi_{1}$ as the variable.%
\footnote{In fact, it is possible to not only exchange the fields but to rotate them by an arbitrary angle $\alpha$.} 
 The discriminant $D$ is invariant under the exchange, but $Q$ and $R$ are not. The allowed parameter space, of course, stays the same. If either $\lambda_{31}$ or $\lambda_{13}$ is zero due to some symmetry, this freedom permits us to simplify the expressions for $Q$ and $R$. For example, if $\lambda_{31} = 0$, the conditions are simpler if we choose $\phi_{1}/\phi_{2}$ as the variable, since the remaining $\lambda_{13}$ term is only linear in $\phi_{1}$. In this case the condition $Q > 0 \lor R > 0$ takes the form $\lambda_{22} + 2 \sqrt{\lambda_{40} \lambda_{04}} > 0$.

As a cross-check, we can set both $\lambda_{31}$ and $\lambda_{13}$ to zero. Then the vacuum stability conditions become
\begin{equation}
\begin{split}
  &\lambda_{40} > 0, \quad \lambda_{04} > 0, \quad D = \lambda_{40} \lambda_{04} 
  (\lambda_{22}^{2} - 4 \lambda_{40} \lambda_{04})^{2} > 0, \\
  & Q = \lambda_{40}^{2} (4 \lambda_{40} \lambda_{04} - \lambda_{22}^{2}) > 0 \lor 
  R = \lambda_{40} \lambda_{22} > 0,
\end{split}
\end{equation}
which can be simplified to
\begin{equation}
  \lambda_{40} > 0, \quad \lambda_{04} > 0, \quad \lambda_{22} 
  + 2 \sqrt{\lambda_{40} \lambda_{04}} > 0,
\label{eq:2:real:2:field:cop}
\end{equation}
the usual copositivity conditions for $V$ with $\phi_{1}^{2}$ and $\phi_{2}^{2}$ as the non-negative variables.

\begin{figure}[tb]
\begin{center}
  \includegraphics{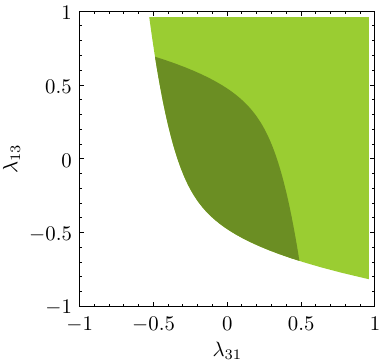}%
  \hspace{0.05\textwidth}
 \includegraphics{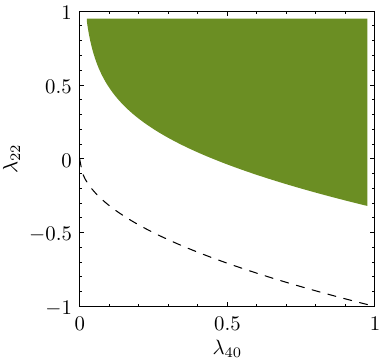}
\caption{Left panel: An example of parameter space allowed by vacuum stability constraints \eqref{eq:2:real:2:field:vac:stab} for the potential \eqref{eq:V:general:2:fields} two real scalars (dark green). If the scalars take only non-negative values, the light green area is allowed in addition \eqref{eq:one:positive:variable:polynomial:positivity}. The values of the remaining parameters are $\lambda_{40} = 0.125$, $\lambda_{04} = 0.25$ and $\lambda_{22} = 0.25$. Right panel: The allowed parameter space (dark green) in the $\lambda_{22}$ vs. $\lambda_{40}$ plane with $\lambda_{04} = 0.25$, $\lambda_{13} = -0.75$ and $\lambda_{31} = 0$. The dashed line in the right panel shows the vacuum stability bound \eqref{eq:2:real:2:field:cop} from copositivity for $\lambda_{31} = \lambda_{13} = 0$.
}
\label{fig:two:real:scalars:lambda13:vs:lambda31}
\end{center}
\end{figure}

An illustration of the vacuum stability conditions for some values of parameters is given in Figure~\ref{fig:two:real:scalars:lambda13:vs:lambda31}. The parts of the parameter space that are allowed are shown in dark green. In addition, the light green area is allowed if the scalars can have only non-negative values (see Section~\ref{sec:two:non-negative:scalars}). If both $\lambda_{31} = \lambda_{13} = 0$, the vacuum stability conditions reduce to the usual copositivity conditions \eqref{eq:2:real:2:field:cop} (dashed line in the right panel).

\subsection{Bounded From Below Conditions for a Quartic Polynomial on $\mathbb{R}_{+}$}
\label{sec:two:non-negative:scalars}

The domain of the general quartic polynomial \eqref{eq:general:quartic} can be restricted to \emph{non-negative} real numbers $\mathbb{R}_{+}$. In this case, the allowed range of parameters is somewhat larger. The positivity conditions for the polynomial \eqref{eq:general:quartic} for $x \geqslant 0$ are given by \cite{Ulrich:1990:PCQ:903209}%
\footnote{We have restored $a \equiv a_{0}$ and $e \equiv a_{4}$ in the conditions given in \cite{Ulrich:1990:PCQ:903209} and slightly reorganised them. They were first used for finding vacuum stability conditions in \cite{Ko:2014nha}.}
\begin{equation}
\begin{split}
  & \left( D \leqslant 0 \land a_{3} \sqrt{a_{0}} + a_{1} \sqrt{a_{4}} > 0 \right) \\
  \lor & 
  \left( -2 \sqrt{a_{0} a_{4}} < a_{2} < 6 \sqrt{a_{0} a_{4}} \land D \geqslant 0 \land \Lambda_{1} \leqslant 0 \right) \\
  \lor &
  \left( 6 \sqrt{a_{0} a_{4}} < a_{2} \land [(a_{1} > 0 \land a_{3} > 0) \lor (D \geqslant 0 
  \land \Lambda_{2} \leqslant 0) ] \right),
\end{split}
\label{eq:one:positive:variable:polynomial:positivity}
\end{equation}
where
\begin{align}
  \Lambda_{1} &= 
  (\sqrt{a_{0}} a_{3} - a_{1} \sqrt{a_{4}})^{2}
   - 32 \left(a_{0} a_{4}\right)^{\frac{3}{2}} - 16 \left(a_{0} a_{2} a_{4}
  + a_{0}^{\frac{5}{4}} a_{3} a_{4}^{\frac{3}{4}} + a_{0}^{\frac{3}{4}} a_{1} a_{4}^{\frac{5}{4}}\right),
  \\
  \Lambda_{2} &=  (\sqrt{a_{0}} a_{3} - a_{1} \sqrt{a_{4}})^{2} - \frac{4 \sqrt{a_{0} a_{4}} (a_{2} + 2 \sqrt{a_{0} a_{4}}) (\sqrt{a_{0}} a_{3} + a_{1} \sqrt{a_{4}} + 4 \sqrt{a_{0} a_{4}} \sqrt{a_{2} - 2 \sqrt{a_{0} a_{4}}})}%
  {\sqrt{ a_{2} - 2 \sqrt{a_{0} a_{4}}}}.
\end{align}

The main difference with the case of two real scalars is that a range of positive and opposite-sign $\lambda_{31}$ and $\lambda_{13}$ is allowed, as seen in the left hand panel of figure~\ref{fig:two:real:scalars:lambda13:vs:lambda31}. Also note that it is trivial to restrict the domain to the negative numbers instead by taking $x \to -x$, equivalent to changing $a_{1} \to -a_{1}$ and $a_{3} \to -a_{3}$ in the above conditions.

If the non-negative variables are magnitudes of scalar fields, then the coefficients $a_{i}$ may depend on additional orbit space parameters, notably phases. These may allow, in effect, to always choose the $\lambda_{31}$ and/or $\lambda_{13}$ terms to be negative. For that reason, as we will see below for the 2HDM, the conditions \eqref{eq:one:positive:variable:polynomial:positivity} for a positive variable can often be eschewed in favour of the simpler conditions \eqref{eq:one:variable:polynomial:positivity} for a real variable.

\subsection{Vacuum Stability Conditions from Positivity with an Affine Space}

We will derive another, different form of vacuum stability conditions for the potential \eqref{eq:V:general:2:fields}. The matrix of quartic couplings of the potential \eqref{eq:V:general:2:fields} in the monomial basis $(\phi_{1}^{2}, \phi_{1} \phi_{2}, \phi_{2}^{2})$ is
\begin{equation}
\displaystyle
  \Lambda = 
  \begin{pmatrix}
  \lambda_{40} & \frac{1}{2} \lambda_{31} & \frac{1}{2} (1-c) \lambda_{22} \\
  \frac{1}{2} \lambda_{31} & c \lambda_{22}  & \frac{1}{2} \lambda_{13} \\
  \frac{1}{2} (1-c) \lambda_{22} & \frac{1}{2} \lambda_{13} & \lambda_{04}
  \end{pmatrix},
\end{equation}
where $c$ is an arbitrary constant due to the ambiguity $(\phi_{1} \phi_{2})^{2} = \phi_{1}^{2} \phi_{2}^{2}$. The matrices $\Lambda (c)$ form an affine space. If for some value of $c$ the matrix $\Lambda$ is positive-definite, then the potential \eqref{eq:V:general:2:fields} is bounded from below. Note that  since $\phi_{1} \phi_{2} \in \mathbb{R}$, one has to demand the usual positivity, not copositivity.

The Sylvester criterion for the positivity of $\Lambda$ is given by
\begin{align}
  \lambda_{40} &> 0, & \lambda_{04} &> 0, & c \lambda_{22} &> 0, 
  \label{eq:2:real:scalars:1:field:conditions} \\
  4 \lambda_{40} \lambda_{04} + (1 - c)^{2} \lambda_{22}^{2} &> 0, 
  & 4 c \lambda_{22} \lambda_{40} - \lambda_{31}^{2} &> 0, 
  & 4 c \lambda_{22} \lambda_{04} - \lambda_{13}^{2} &> 0,
  \label{eq:2:real:scalars:2:field:conditions}
  \\
  \span \span \span \span c \lambda_{22} (4 \lambda_{40} \lambda_{04} - \lambda_{31} \lambda_{13})
  - \lambda_{40} \lambda_{13}^{2} 
  + \lambda_{22} \lambda_{31} \lambda_{13}
  - \lambda_{04} \lambda_{31}^{2}
  - c (1 - c)^{2} \lambda_{22}^{3}  &> 0.
\label{eq:2:real:scalars:3:field:conditions}
\end{align}

First of all, the last inequality of \eqref{eq:2:real:scalars:1:field:conditions} implies that $c$ is real and has the same sign as $\lambda_{22}$. The first inequality of \eqref{eq:2:real:scalars:2:field:conditions} is trivially satisfied. To satisfy the last two inequalities of \eqref{eq:2:real:scalars:2:field:conditions} one must have $\abs{c} \geqslant \abs{c_{0}}$, where
\begin{equation}
  c_{0} \equiv  \frac{1}{4 \lambda_{22}} \max \left( \frac{\lambda_{13}^{2}}{\lambda_{40}}, 
  \frac{\lambda_{31}^{2}}{\lambda_{04}} \right).
\end{equation}
The extrema of the left hand side (LHS) of the inequality \eqref{eq:2:real:scalars:3:field:conditions} with respect to $c$ are
\begin{equation}
  c_{\pm} = \frac{1}{3} \left[ 2 \pm \frac{\sqrt{\lambda_{22}^{2} (\lambda_{22}^{2} + 12 \lambda_{40} \lambda_{04} - 3 \lambda_{31} \lambda_{13})}}{\lambda_{22}^{2}} \right].
\end{equation}
Which of $c_{\pm}$ gives the maximum of the LHS of \eqref{eq:2:real:scalars:3:field:conditions}? The second derivative of the LHS of \eqref{eq:2:real:scalars:3:field:conditions} with respect to $c$ is $(1- \frac{3}{2} c) \lambda_{22}^{3}$. Inserting here the solutions $c_{\pm}$, we see that for $\lambda_{22} > 0$, the maximum is given by $c_{+}$ and likewise for $\lambda_{22} < 0$ the maximum is $c_{-}$. Of course, if for $\lambda_{22} > 0$, for example, $c_{+} < c_{0}$, one must take $c = c_{0}$.

Therefore, the optimal value for $c$ is
\begin{equation}
  c = 
  \begin{cases}
    \max (c_{0}, c_{+}) & \text{if } \lambda_{22} > 0, \\
    \min (c_{0}, c_{-}) & \text{if } \lambda_{22} < 0, \\
    c = 0            & \text{if } \lambda_{22} = 0.
  \end{cases}
  \label{eq:c:2:real:3:field:conditions:gen}
\end{equation}

The conditions \eqref{eq:2:real:scalars:1:field:conditions}, \eqref{eq:2:real:scalars:2:field:conditions} and \eqref{eq:2:real:scalars:3:field:conditions} together with \eqref{eq:c:2:real:3:field:conditions:gen} define the same region of the parameter space as \eqref{eq:2:real:2:field:vac:stab}. We have traded the relative complexity of the latter for the apparent simplicity of the former at the cost of introducing the optimal coefficient $c$ as a piece-wise function of the couplings that has a discontinuity at $\lambda_{22} = 0$.

\section{Vacuum Stability for Two Real Scalars \emph{\&} the Higgs boson}
\label{sec:two:real:scalars:and:Higgs}

The most general scalar potential of two real scalar fields $\phi_{1}$ and $\phi_{2}$ and the Higgs doublet $H$ is
\begin{equation}
\begin{split}
  V(\phi_{1}, \phi_{2}, \abs{H}^{2}) &= \lambda_{H} \abs{H}^{4} 
  +  \lambda_{H20} \abs{H}^{2} \phi_{1}^{2} + \lambda_{H11} \abs{H}^{2} \phi_{1} \phi_{2} 
  + \lambda_{H02} \abs{H}^{2} \phi_{2}^{2}
  \\
  & + \lambda_{40} \phi_{1}^{4} + \lambda_{31} \phi_{1}^{3} \phi_{2} 
  + \lambda_{22} \phi_{1}^{2} \phi_{2}^{2} + \lambda_{13} \phi_{1} \phi_{2}^{3} 
  + \lambda_{04} \phi_{2}^{4} \\
  &\equiv \lambda_{H} \abs{H}^{4} + M^{2}(\phi_{1}, \phi_{2}) \abs{H}^{2} + V(\phi_{1}, \phi_{2}),
\end{split}
\label{eq:V:general:2:fields:H}
\end{equation}
where $M^{2}(\phi_{1}, \phi_{2}) \equiv \lambda_{H20} \phi_{1}^{2} + \lambda_{H11} \phi_{1} \phi_{2} + \lambda_{H02} \phi_{2}^{2}$ and $V(\phi_{1}, \phi_{2}) \equiv V(\phi_{1}, \phi_{2}, 0)$.

The potential \eqref{eq:V:general:2:fields:H} is a quadratic polynomial in $\abs{H}^{2}$. Setting $\phi_{1} = \phi_{2} = 0$, we obtain $\lambda_{H} > 0$. Setting $\abs{H}^{2} = 0$ we recover the conditions \eqref{eq:2:real:2:field:vac:stab} for $V(\phi_{1}, \phi_{2}) > 0$. If all three fields are non-zero, we can eliminate the Higgs: $V(\phi_{1}, \phi_{2}, \abs{H}^{2}) > 0$ if either $M^{2}(\phi_{1}, \phi_{2}) > 0$, or else the discriminant of $V(\phi_{1}, \phi_{2}, \abs{H}^{2}) $ with respect to $\abs{H}^{2}$ is negative, that is $M^{4}(\phi_{1}, \phi_{2}) - 4 \lambda_{H} V(\phi_{1}, \phi_{2}) < 0$.

An equivalent way to eliminate $\abs{H}^{2}$ is to minimise the potential with respect to it:
\begin{equation}
  0 = \frac{\partial V}{\partial \abs{H}^{2}} = 2 \lambda_{H} \abs{H}^{2} + M^{2}(\phi_{1}, \phi_{2}),
\end{equation}
giving
\begin{equation}
  \abs{H}^{2}_{\rm min} = -\frac{1}{2 \lambda_{H}} M^{2}(\phi_{1}, \phi_{2}).
  \label{eq:H:min}
\end{equation}
Again, either $M^{2}(\phi_{1}, \phi_{2}) > 0$ and the solution for $\abs{H}^{2}_{\rm min}$ is unphysical, or else $V_{\abs{H}^{2} = \abs{H}^{2}_{\rm min}} = V(\phi_{1}, \phi_{2}) - \frac{1}{4 \lambda_{H}} M^{4}(\phi_{1}, \phi_{2})$ must be positive. Inserting the solution \eqref{eq:H:min} into the potential in effect means substituting $\lambda_{40} \to \lambda_{40} - \lambda_{H20}^{2}/\lambda_{H}$ and so on in $V(\phi_{1}, \phi_{2})$.

Therefore, for the potential \eqref{eq:V:general:2:fields:H} to be bounded from below, we altogether require 
\begin{equation}
 \! \! \! \lambda_{H} > 0,  \;\;\: V(\phi_{1}, \phi_{2}) > 0, \;\;\: 
  M^{2}(\phi_{1}, \phi_{2}) > 0 \; \lor \; V_{\abs{H}^{2} = \abs{H}^{2}_{\rm min}}(\phi_{1}, \phi_{2}) =  V(\phi_{1}, \phi_{2}) - \frac{1}{4 \lambda_{H}} M^{4}(\phi_{1}, \phi_{2}) > 0,
\end{equation}
where the last \textsc{or} condition is tantamount to the problem of positivity of the quartic polynomial $V_{\abs{H}^{2} = \abs{H}^{2}_{\rm min}}$ with the quadratic constraint $M^{2} < 0$. In general, $M^{2}(\phi_{1}, \phi_{2})$ can be positive for some values of $\phi_{1}$ and $\phi_{2}$ and negative for others. In this case the region defined by $M^{2}(\phi_{1}, \phi_{2}) < 0$ is a pointed double cone in the $\phi_{1} \phi_{2}$-plane. 

In some regions of the parameter space it is easy to find the conditions. If the coefficient matrix of $M^{2}$, given by
\begin{equation}
  \mathbf{M}^{2} = 
  \begin{pmatrix}
    \lambda_{H20} & \frac{1}{2} \lambda_{H11} \\
    \frac{1}{2} \lambda_{H11} & \lambda_{H02}
  \end{pmatrix},
\end{equation}
is positive-definite, that is
\begin{equation}
  \lambda_{H20} > 0, \quad \lambda_{H02} > 0, \quad 4 \lambda_{H20} \lambda_{H02} > \lambda_{H11}^{2},
  \label{eq:positive:M:2:conditions}
\end{equation}
then $M^{2} > 0$ for any values of the fields. If, on the other hand,
\begin{equation}
  \lambda_{H20} \leqslant 0, \quad \lambda_{H02} \leqslant 0, \quad 4 \lambda_{H20} \lambda_{H02} \leqslant \lambda_{H11}^{2},
  \label{eq:negative:M:2:conditions}
\end{equation}
then $M^{2} \leqslant 0$ for any values of the fields, and $V_{\abs{H}^{2} = \abs{H}^{2}_{\rm min}} > 0$ must hold for all values of the fields. 

The intermediate situation, where for some values of the fields $M^{2} < 0$ and for others $M^{2} > 0$, occurs if the eigenvalues of $\mathbf{M}^{2}$ have opposite sign, that is, the determinant of $\mathbf{M}^{2}$ is negative:
\begin{equation}
  4 \lambda_{H20} \lambda_{H02} < \lambda_{H11}^{2}.
\end{equation}
Then we can always make a transformation of the singlet fields to bring $M^{2}$ into the anti-diagonal form 
\begin{equation}
  M^{\prime 2} = \lambda'_{H11} \phi_{1} \phi_{2}.
  \label{eq:M:2:antidiag}
\end{equation}
It is evident that such an $M^{\prime 2}$ is negative in two opposite quadrants in the $\phi_{1}\phi_{2}$-plane. If we dehomogenise $V_{\abs{H}^{2} = \abs{H}^{2}_{\rm min}}$ by taking e.g. $\phi_{1} = 1$, we have to restrict the other field $\phi_{2}$ to a half-axis to respect the constraint $M^{\prime 2} < 0$. But we already have conditions for a quartic polynomial to be positive (or negative) on a half-axis: the conditions \eqref{eq:one:positive:variable:polynomial:positivity} for a quartic in a positive variable.

To begin to bring $M^{2}$ into the form \eqref{eq:M:2:antidiag}, first we diagonalise $\mathbf{M}^{2}$ by the orthogonal matrix
\begin{equation}
  \mathbf{U}_{\theta} = 
  \begin{pmatrix}
    \cos \theta & -\sin \theta
    \\
    \sin \theta & \phantom{-} \cos \theta
  \end{pmatrix},
\end{equation}
where
\begin{align}
  \sin \theta &= \frac{\lambda_{H11} \sgn (\lambda_{H02} - \lambda_{H20})}{\sqrt{\lambda_{H11}^{2} +\left[ \sqrt{(\lambda_{H20}-\lambda_{H02})^{2}} + \sqrt{(\lambda_{H20}-\lambda_{H02})^{2} +  \lambda_{H11}^{2}} \right]^{2}}},
  \\
  \cos \theta &= \frac{\sqrt{(\lambda_{H20}-\lambda_{H02})^{2}} + \sqrt{(\lambda_{H20}-\lambda_{H02})^{2} +  \lambda_{H11}^{2}}}{\sqrt{\lambda_{H11}^{2} +\left[ \sqrt{(\lambda_{H20}-\lambda_{H02})^{2}} + \sqrt{(\lambda_{H20}-\lambda_{H02})^{2} +  \lambda_{H11}^{2}} \right]^{2}}}.
\end{align}
For our purposes, $\sgn 0 = 1$.
The eigenvalues of $\mathbf{M}^{2}$ are given by
\begin{equation}
  \lambda'_{H\mp} = \frac{1}{2} \left[ \lambda_{H20} + \lambda_{H02} \mp \sqrt{\lambda_{H11}^{2}
  + (\lambda_{H20} - \lambda_{H02})^{2}} \right].
\end{equation}
In the intermediate case, we have $\lambda'_{H-} < 0$ and $\lambda'_{H+} > 0$.
After the diagonalisation
\begin{equation}
  \mathbf{M}^{2} = 
  \begin{pmatrix}
    \lambda'_{H20} & 0
    \\
    0 & \lambda'_{H02}
  \end{pmatrix}
\end{equation}
with $\abs{\lambda'_{H20}} = -\lambda'_{H-},~ \abs{\lambda'_{H02}} = \lambda'_{H+}$ if $\lambda_{H20} < \lambda_{H02}$, and $\abs{\lambda'_{H20}} = \lambda'_{H+},~ \abs{\lambda'_{H02}} = -\lambda'_{H-}$ if $\lambda_{H20} \geqslant \lambda_{H02}$, so
\begin{align}
  \abs{\lambda'_{H20, H02}} &= \mp\frac{1}{2} \sgn (\lambda_{H02} - \lambda_{H20}) \left[ \lambda_{H20} + \lambda_{H02} \mp \sgn (\lambda_{H02} - \lambda_{H20}) \sqrt{\lambda_{H11}^{2}
  + (\lambda_{H20} - \lambda_{H02})^{2}} \right].
\end{align}
Next we scale $\mathbf{M}^{2}$ by 
\begin{equation}
  \mathbf{S} = 
  \frac{1}{\sqrt{2}}
  \begin{pmatrix}
    \frac{1}{\sqrt{\abs{\lambda'_{H20}}}} & 0 \\
    0 & \frac{1}{\sqrt{\abs{\lambda'_{H02}}}}
  \end{pmatrix}
\end{equation}
to make it proportional to the unit matrix.
Finally we rotate $\mathbf{M}^{2}$ by
\begin{equation}
  \mathbf{U}_{\frac{\pi}{4}} = 
  \frac{1}{\sqrt{2}}
  \begin{pmatrix}
    1 & -1 \\
    1 & \phantom{-}1
  \end{pmatrix}
\end{equation}
into the anti-diagonal form.
Altogether, we transform the fields in $V_{\abs{H}^{2} = \abs{H}^{2}_{\rm min}}$ by 
\begin{equation}
  \begin{pmatrix}
    \phi_{1} \\ \phi_{2}
  \end{pmatrix}
  \to
  \T{\mathbf{U}_{\theta}}\mathbf{S} \,\mathbf{U}_{\frac{\pi}{4}}   
  \begin{pmatrix}
    \phi_{1} \\ \phi_{2}
  \end{pmatrix},
\end{equation}
yielding the transformed potential $V'_{\abs{H}^{2} = \abs{H}^{2}_{\rm min}}$.

\begin{figure}[tb]
\begin{center}
\includegraphics{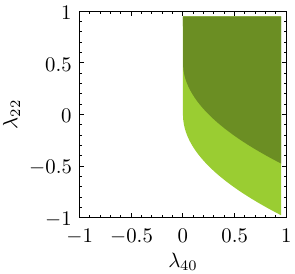}
\includegraphics{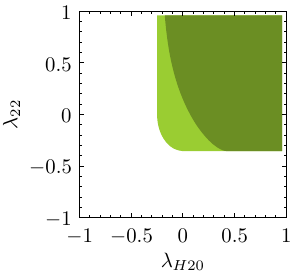}
\includegraphics{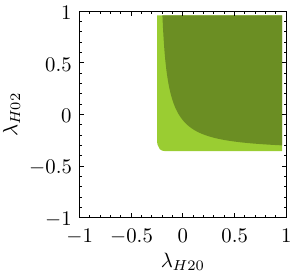}
\\
\vspace{4mm}
\includegraphics{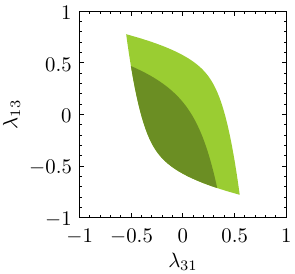}
\includegraphics{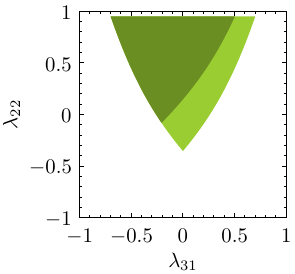}
\includegraphics{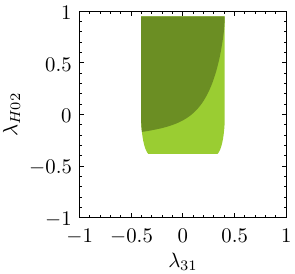}
\caption{An example of parameter space allowed by the vacuum stability conditions \eqref{eq:V:general:2:fields:H:vac:stab} for the potential \eqref{eq:V:general:2:fields:H} of two real scalars and the Higgs. The values of parameters are $\lambda_{H} = 0.125$, $\lambda_{40} = 0.125$, $\lambda_{04} = 0.25$, $\lambda_{31} = \lambda_{13} = \lambda_{H20} = \lambda_{H02} = 0$, except for plots where these couplings vary. Both light and dark green regions are allowed for for $\lambda_{H11} = 0$, while only the dark green region is allowed for $\lambda_{H11} = 0.5$.}
\label{fig:12H:plots}
\end{center}
\end{figure}

After the transformation, the coefficient matrix has the form
\begin{equation}
  \mathbf{M}^{\prime 2} = 
  \frac{1}{2} \sgn (\lambda_{H02} - \lambda_{H20})
  \begin{pmatrix}
    0 & 1 \\
    1 & 0
  \end{pmatrix},
\end{equation}
hence $\lambda'_{H11} = \sgn (\lambda_{H02} - \lambda_{H20})$. 

 We can now dehomogenise $V'_{\abs{H}^{2} = \abs{H}^{2}_{\rm min}}$ by taking $\phi_{1} = 1$, and use the conditions \eqref{eq:one:positive:variable:polynomial:positivity} for when $M^{\prime 2}$ is negative. If $\lambda'_{H11} < 0$, then $M^{\prime 2}$ is negative in the 1st and 3rd quadrants of the $\phi_{1}\phi_{2}$-plane and the conditions \eqref{eq:one:positive:variable:polynomial:positivity} apply as is; if $\lambda'_{H11} > 0$, then $M^{\prime 2}$ is negative in the 2nd and 4th quadrants, and we have to take $\phi_{2} \to -\phi_{2}$ in $V'_{\abs{H}^{2} = \abs{H}^{2}_{\rm min}}$ before applying the conditions. In short, we have to take $\phi_{2} \to -\sgn (\lambda'_{H11}) \, \phi_{2}$, equivalent to taking $\lambda'_{31} \to -\sgn (\lambda'_{H11}) \, \lambda'_{31}$ and $\lambda'_{13} \to -\sgn (\lambda'_{H11}) \, \lambda'_{13}$.

Altogether, the vacuum stability conditions for the potential \eqref{eq:V:general:2:fields:H} are given by
\begin{equation}
\begin{split}
  & \lambda_{40} > 0, \quad \lambda_{04} > 0, \quad \lambda_{H} > 0, \quad
  D_{\abs{H}^{2} = 0} \land (Q_{\abs{H}^{2} = 0} > 0 \lor R_{\abs{H}^{2} = 0} > 0),
  \\
  & \lambda_{H20} \leqslant 0 \land \lambda_{H02} \leqslant 0 
  \land \lambda_{H11}^{2} \leqslant 4 \lambda_{H20} \lambda_{H02}
  \implies 
  4 \lambda_{H} \lambda_{40} - \lambda_{H20}^{2} > 0 
  \land 4 \lambda_{H} \lambda_{04} - \lambda_{H02}^{2} > 0 
  \\
  & \land D_{\abs{H}^{2} = \abs{H}^{2}_{\rm min}} \land 
  (Q_{\abs{H}^{2} = \abs{H}^{2}_{\rm min}} > 0 \lor R_{\abs{H}^{2} 
  = \abs{H}^{2}_{\rm min}} > 0),
  \\
  & \lambda_{H11}^{2} > 4 \lambda_{H20} \lambda_{H02} 
  \implies
  \lambda'_{04} > 0 \land \lambda'_{40} > 0
  \\
  &\land \left[ \left( D_{\abs{H}^{2} = \abs{H}^{2}_{\rm min}} \leqslant 0 \land 
  \left( \lambda'_{31} \sqrt{\lambda'_{04}} + \lambda'_{13} \sqrt{\lambda'_{40}} \right) > 0 \right) \right. \\
  &\lor 
  \left( -2 \sqrt{\lambda'_{04} \lambda'_{40}} < \lambda'_{22} < 6 \sqrt{\lambda'_{04} \lambda'_{40}} \land D_{\abs{H}^{2} = \abs{H}^{2}_{\rm min}} \geqslant 0 
  \land \Lambda'_{1 \: \abs{H}^{2} = \abs{H}^{2}_{\rm min}} \leqslant 0 \right) 
  \\
  &\lor 
  \left( 6 \sqrt{\lambda'_{04} \lambda'_{40}} < \lambda'_{22} \land 
  [(\lambda'_{13} > 0 \land \lambda'_{31} > 0) 
 \left. \lor (D_{\abs{H}^{2} = \abs{H}^{2}_{\rm min}} \geqslant 0 
  \land \Lambda'_{2  \: \abs{H}^{2} = \abs{H}^{2}_{\rm min}} \leqslant 0) ] \right) \right],
\end{split}
\label{eq:V:general:2:fields:H:vac:stab}
\end{equation}
where we take $\lambda'_{31} \to -\sgn (\lambda_{H02} - \lambda_{H20}) \lambda'_{31}$ and $\lambda'_{13} \to -\sgn (\lambda_{H02} - \lambda_{H20}) \lambda'_{13}$. The primes on $\Lambda'_{1}$ and $\Lambda'_{2}$ indicate that they are calculated for the transformed potential $V'_{\abs{H}^{2} = \abs{H}^{2}_{\rm min}}$. The discriminant of the transformed potential is, up to a positive constant coefficient which does not affect positivity, equal to the discriminant of $V_{\abs{H}^{2} = \abs{H}^{2}_{\rm min}}$ and does not depend on $\sgn (\lambda_{H02} - \lambda_{H20})$. 
The conditions \eqref{eq:V:general:2:fields:H:vac:stab} are illustrated in Figure~\ref{fig:12H:plots}.

For comparison, if we set all the terms containing odd powers of $\phi_{1}$ and $\phi_{2}$ to zero, $\lambda_{H11} = \lambda_{31} = \lambda_{13} = 0$, then the positivity of the potential is given by the much simpler (strict) copositivity constraints on the matrix of couplings in the $(\phi_{1}, \phi_{2}, \abs{H}^{2})$ basis,
\begin{align}
  \lambda_{40} &> 0, &  \lambda_{04} &> 0, & \lambda_{H} &> 0,
  \notag
  \\
  \bar{\lambda}_{22} \equiv \lambda_{22} + 2 \sqrt{\lambda_{40} \lambda_{04}} &> 0,
  &
  \bar{\lambda}_{H20} \equiv \lambda_{H20} + 2 \sqrt{\lambda_{40} \lambda_{H}} &> 0,
  &
  \bar{\lambda}_{H02} \equiv \lambda_{H02} + 2 \sqrt{\lambda_{04} \lambda_{H}} &> 0,
  \\
  \span \span  \span \span \sqrt{\lambda_{40}} \lambda_{H02} + \sqrt{\lambda_{04}} \lambda_{H20} + \sqrt{\lambda_{H}} \lambda_{22} 
  + \sqrt{\lambda_{40} \lambda_{04} \lambda_{H}} + \sqrt{\bar{\lambda}_{22} \bar{\lambda}_{H20} \bar{\lambda}_{H02}} &> 0,
  \notag
\end{align}
which are still necessary, but not sufficient conditions for positivity in the general case.

The transformed potential $V'_{\abs{H}^{2} = \abs{H}^{2}_{\rm min}}$ is rather complicated, so let us look at special cases. If $\lambda_{H20} = \lambda_{H02} = 0$, then $\mathbf{M}^{2}$ already has the anti-diagonal form, $V'_{\abs{H}^{2} = \abs{H}^{2}_{\rm min}} = V_{\abs{H}^{2} = \abs{H}^{2}_{\rm min}}$ and $\lambda'_{H11} = \lambda_{H11}$.

If, on the other hand, $\lambda_{H11} = 0$, then $\mathbf{M}^{2}$ is diagonal and $U_{\theta}$ is the unit matrix. Then $\lambda_{H20}$ and $\lambda_{H02}$ must have opposite signs for the determinant of $\mathbf{M}^{2}$ to be negative and $\sgn (\lambda_{H02} - \lambda_{H20}) = \sgn \lambda_{H02}$. The transformed potential (multiplied by the irrelevant positive coefficient $4 \lambda_{H20}^{2} \lambda_{H02}^{2}$ to make it simpler) is 
\begin{equation}
  4 \lambda_{H20}^{2} \lambda_{H02}^{2} 
  V'_{\abs{H}^{2} = \abs{H}^{2}_{\rm min}} = \lambda'_{40} \phi_{1}^{4} + \lambda'_{31} \phi_{1}^{3} \phi_{2} + \lambda'_{22} \phi_{1}^{2} \phi_{2}^{2} + \lambda'_{13} \phi_{1} \phi_{2}^{3} + \lambda'_{04} \phi_{2}^{4},
\end{equation}
where 
\begin{equation}
\begin{split}
  \lambda'_{40} &= \lambda_{H} [ \lambda_{H20}^2 \lambda_{04} - \lambda_{22} 
  \lambda_{H20} \lambda_{H02} + \lambda_{40} \lambda_{H02}^2 
  + \sqrt{-\lambda_{H20} \lambda_{H02}} (\lambda_{13} \abs{\lambda_{H20}} + 
  \lambda_{31} \abs{\lambda_{H02}}) ],
  \\
  \lambda'_{31} &= 2 \lambda_{H} [ 2 \lambda_{H20}^2 \lambda_{04} - 2 \lambda_{40} 
  \lambda_{H02}^2 + \sqrt{-\lambda_{H20} \lambda_{H02}} (\lambda_{13} \abs{\lambda_{H20}} - 
  \lambda_{31} \abs{\lambda_{H02}}) ],
  \\
  \lambda'_{22} &= 2\lambda_{H} (3 \lambda_{40} \lambda_{H02}^2 + \lambda_{22} \lambda_{H20} \lambda_{H02} + 3 \lambda_{04} \lambda_{H20}^2)
  - 4 \lambda_{H20}^2 \lambda_{H02}^2,
  \\
  \lambda'_{13} &= 2 \lambda_{H}  [ 2 \lambda_{H20}^2 \lambda_{04} - 2 \lambda_{40} \lambda_{H02}^2 
  - \sqrt{-\lambda_{H20} \lambda_{H02}} (\lambda_{13} \abs{\lambda_{H20}} 
  - \lambda_{31} \abs{\lambda_{H02}}) ],
  \\
  \lambda'_{04} &= \lambda_{H} [ \lambda_{H20}^2 \lambda_{04} - \lambda_{22} \lambda_{H20} \lambda_{H02} + \lambda_{40} \lambda_{H02}^2 - \sqrt{-\lambda_{H20} \lambda_{H02}} (\lambda_{13} \abs{\lambda_{H20}} + 
\lambda_{31} \abs{\lambda_{H02}}) ].
\end{split}
\label{eq:V:1:2:H:min:H:transformed}
\end{equation}
In several particle physics models, the equivalent of $M^{2}$ already is diagonal. An example is given by the classically scale invariant $\mathbb{Z}_{3}$ symmetric dark matter model \cite{Ko:2014nha}.

\section{Vacuum Stability of the 2HDM with Real Couplings}
\label{sec:2hdm}

The scalar potential of two Higgs doublets $H_{1}$ and $H_{2}$ in the 2HDM with no explicit CP-violation is
\begin{equation}
\begin{split}
  V &= \lambda_{1} \abs{H_{1}}^{4} + \lambda_{2} \abs{H_{2}}^{4} 
  + \lambda_{3} \abs{H_{1}}^{2} \abs{H_{2}}^{2} + \lambda_{4} (\hc{H_{1}} H_{2}) (\hc{H_{2}} H_{1}) 
  + \frac{1}{2} \lambda_{5} \left[  (\hc{H_{1}} H_{2})^{2} + (\hc{H_{2}} H_{1})^{2} \right] \\
  &+ \lambda_{6} \abs{H_{1}}^{2} ( \hc{H_{1}} H_{2} + \hc{H_{2}} H_{1})
  + \lambda_{7} \abs{H_{2}}^{2} (\hc{H_{1}} H_{2} + \hc{H_{2}} H_{1})
   \\
  & = \lambda_{1} h_{1}^{4} + \lambda_{2} h_{2}^{4} 
  + \lambda_{3} h_{1}^{2} h_{2}^{2} + \lambda_{4} \rho^{2} h_{1}^{2} h_{2}^{2}
  + \lambda_{5} \rho^{2} \cos 2 \phi \: h_{1}^{2} h_{2}^{2} + 2 \lambda_{6} \rho \cos \phi \: h_{1}^{3} h_{2} 
  + 2  \lambda_{7} \rho \cos \phi \: h_{1} h_{2}^{3},
\end{split}
\label{eq:V:2HDM}
\end{equation}
where we have taken the potentially complex couplings $\lambda_{5}$, $\lambda_{6}$ and $\lambda_{7}$ real and parameterised the field bilinears as \cite{Ginzburg:2004vp}
\begin{equation}
  \abs{H_{1}}^{2} = h_{1}^{2}, \quad \abs{H_{2}}^{2} = h_{2}^{2}.
  \quad \hc{H_{1}} H_{2} = h_{1} h_{2} \rho e^{i \phi}.
  \label{eq:V:2HDM:param}
\end{equation}
The orbit space parameter $\rho \in [0,1]$ as implied by the Cauchy inequality $0 \leqslant \abs{\hc{H_{1}} H_{2}} \leqslant \abs{H_{1}} \abs{H_{2}}$. 
While the general form of the vacuum stability conditions for the most general 2HDM potential has been given \cite{Maniatis:2006fs,Ivanov:2006yq,Ivanov:2015nea} in the elegant `light cone' formalism, the conditions can be given in a simple explicit and analytical form in terms of the potential couplings only in special cases.

If $\lambda_{6} = \lambda_{7} = 0$, we recover the vacuum stability conditions \cite{Deshpande:1977rw,Klimenko:1984qx,Nie:1998yn,Kanemura:1999xf,Ginzburg:2004vp} for the inert doublet model (IDM)
\begin{gather}
  \lambda_{1} > 0,  \quad \lambda_{2} > 0, \quad \lambda_{3} + 2 \sqrt{\lambda_{1} \lambda_{2}} > 0, 
  \label{eq:ID:copos:crit:first}
  \\
  \lambda_{3} + \lambda_{4} - \abs{\lambda_{5}} + 2 \sqrt{\lambda_{1} \lambda_{2}} > 0.
  \label{eq:ID:copos:crit:last}
\end{gather}
If $\lambda_{6} \neq 0$ or $\lambda_{7} \neq 0$, then the condition $\lambda_{3} + \lambda_{4} - \lambda_{5} + 2 \sqrt{\lambda_{1} \lambda_{2}} > 0$ is a necessary condition \cite{Eriksson:2009ws}.

Finding the minimum of the general potential \eqref{eq:V:2HDM} is complicated because of its non-linear dependence on orbit parameters $\rho$ and $\phi$ \cite{Kim:1982dn}. It is practically impossible to minimise the polynomials $D$, $\Lambda_{1}$ and $\Lambda_{2}$ in the conditions \eqref{eq:one:positive:variable:polynomial:positivity} with respect to these parameters. Instead, we minimise the potential with respect to $\phi$, $\rho$, $h_{1}$, and $h_{2}$,  with the fields lying on the circle $h_{1}^{2} + h_{2}^{2} = 1$, enforced by a Lagrange multiplier $\lambda$ as in eq. \eqref{eq:V:Lagrange:multiplier}. The minimisation equations are
\begin{align}
  h_{1} h_{2} \rho \left( 2 \lambda_{5} \rho h_{1} h_{2} \cos \phi + \lambda_{6} h_{1}^{2} + \lambda_{7} h_{2}^{2} \right) \sin \phi &=0,
  \label{eq:2HDM:min:phi}
  \\
  h_{1} h_{2} \left[ \left(\lambda_{4} + \lambda_{5} \cos 2 \phi \right) 
  \rho h_{1} h_{2} + \left(\lambda_{6} h_{1}^{2} + \lambda_{7} h_{2}^{2} \right) \cos \phi \right] &= 0,
  \label{eq:2HDM:min:rho}
  \\
  4 \lambda_{1} h_{1}^{3} +2 [\lambda_{3} + \left(\lambda_{4} 
  + \lambda_{5} \cos 2 \phi \right) \rho^{2}] \, h_{1} h_{2}^{2}
  + 6 \lambda_{6} \rho \cos \phi \: h_{1}^{2} h_{2}
  + 2 \lambda_{7} \rho \cos \phi \: h_{2}^{3} &= \lambda h_{1}, 
  \label{eq:2HDM:min:h1}
  \\
   4 \lambda_{2} h_{2}^{3} + 2 [\lambda_{3} + \left(\lambda_{4} 
   + \lambda_{5} \cos 2 \phi \right) \rho^{2}] h_{1}^{2} h_{2}
  + 2 \lambda_{6} \rho \cos \phi \: h_{1}^{3}
  + 6 \lambda_{7} \rho \cos \phi \: h_{1} h_{2}^{2} &= \lambda h_{2},
  \label{eq:2HDM:min:h2}
  \\
  h_{1}^{2} + h_{2}^{2} &= 1.
  \label{eq:2HDM:min:lambda}
\end{align}

\begin{figure}[tb]
\begin{center}
\includegraphics{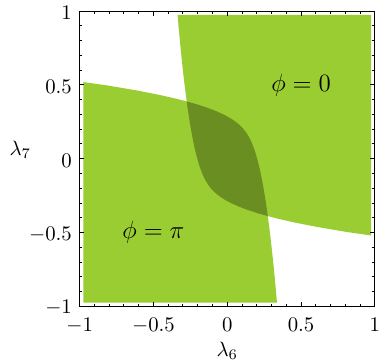}
\caption{The allowed region in the $\lambda_{7}$ vs. $\lambda_{6}$ plane for the 2HDM with no CP-violation is the intersection (dark green) of the regions with $\phi = 0$ and $\phi = \pi$ (light green). Other parameters have values $\lambda_{1} = 0.125$, $\lambda_{2} = 0.25$ and $\lambda_{3} = \lambda_{4} = \lambda_{5} = 0$.}
\label{fig:2HDM:no:CPV:lambda:7:vs:lambda:6}
\end{center}
\end{figure}

Eq. \eqref{eq:2HDM:min:phi} reduces to $\sin \phi = 0$ which yields $\phi = 0$ and $\phi = \pi$ or $\cos \phi = \pm 1$. (There is another solution for $\phi$ that holds in the special case $\lambda_{4} = \lambda_{5}$, but in this limit it gives the same solution for $V_{\text{min}}$ as we will obtain below.) The solutions for $h_{1} = 0$, $h_{2} = 0$ and $\rho = 0$ reproduce the conditions \eqref{eq:ID:copos:crit:first}. While each of the two solutions gives a region of parameter space corresponding to the allowed region \eqref{eq:one:positive:variable:polynomial:positivity} for a positive variable, the allowed region is their \emph{intersection} which is given by the allowed region \eqref{eq:one:variable:polynomial:positivity} of parameter space for a \emph{real} variable for $V_{\phi = 0}$ \emph{or} $V_{\phi = \pi}$, illustrated in figure~\ref{fig:2HDM:no:CPV:lambda:7:vs:lambda:6}.

The extremum solutions to the equations \eqref{eq:2HDM:min:rho}, \eqref{eq:2HDM:min:h1}, \eqref{eq:2HDM:min:h2} and \eqref{eq:2HDM:min:lambda} are given by
\begin{align}
  \rho^{2} &= \frac{(\lambda_{6} h_{1}^{2} + \lambda_{7} h_{2}^{2})^{2}}{h_{1}^{2} 
  h_{2}^{2} (\lambda_{4} + \lambda_{5})^{2}}, \label{eq:2HDM:noCPV:rho}
  \\
  h_{1}^{2} &= \frac{1}{2} \frac{(2 \lambda_{2} - \lambda_{3}) 
  (\lambda_{4} + \lambda_{5}) + 2 \lambda_{7} (\lambda_{6} - \lambda_{7}) }
  {(\lambda_{1} + \lambda_{2} - \lambda_{3}) (\lambda_{4} + \lambda_{5}) 
  - (\lambda_{6} - \lambda_{7})^{2}},
\label{eq:2HDM:noCPV:h1}
  \\
  h_{2}^{2} &= \frac{1}{2} \frac{(2 \lambda_{1} - \lambda_{3}) 
  (\lambda_{4} + \lambda_{5}) - 2 \lambda_{6} (\lambda_{6} - \lambda_{7}) }
  {(\lambda_{1} + \lambda_{2} - \lambda_{3}) (\lambda_{4} + \lambda_{5}) 
  - (\lambda_{6} - \lambda_{7})^{2}},
\label{eq:2HDM:noCPV:h2}
  \\
  V_{\text{min}} &=  \frac{1}{4} \frac{(\lambda_{4} + \lambda_{5}) 
  (4 \lambda_{1} \lambda_{2} - \lambda_{3}^{2}) - 4 (\lambda_{1} \lambda_{7}^{2} 
  + \lambda_{2} \lambda_{6}^{2} - \lambda_{3} \lambda_{6} \lambda_{7})
  }%
  {(\lambda_{1} + \lambda_{2} - \lambda_{3}) (\lambda_{4} + \lambda_{5}) 
  - (\lambda_{6} - \lambda_{7})^{2}}.
\label{eq:2HDM:noCPV:V}
\end{align}
Note that the solution for $\rho^{2}$ in \eqref{eq:2HDM:noCPV:rho} is non-negative if $h_{1}^{2}$ and $h_{2}^{2}$ are  non-negative.

\begin{figure}[tb]
\begin{center}
\includegraphics{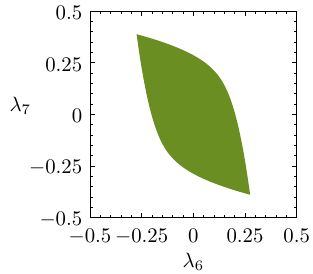}~\includegraphics{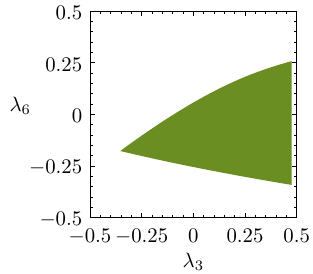}~\includegraphics{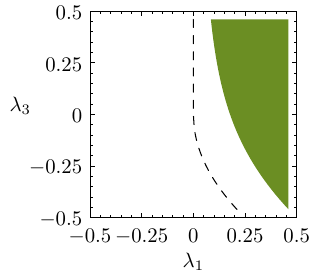}
\caption{The allowed parameter space for some values of the couplings of the 2HDM.  For all panels $\lambda_{2} = 0.25$, $\lambda_{4} = 0.25$, $\lambda_{5} = 0$. The values of other parameters are $\lambda_{1} = 0.125$, $\lambda_{3} = 0$ for the first panel, $\lambda_{1} = 0.125$, $\lambda_{7} = 0.25$ for the second panel, and $\lambda_{6} = 0.25$, $\lambda_{7} = 0$ for the third panel (the dashed line shows the bound from copositivity for $\lambda_{6} = \lambda_{7} = 0$).}
\label{fig:2HDM:parameter:space}
\end{center}
\end{figure}

Because the extremum may not be a minimum or may lie outside the rectangular orbit space $\rho \in [0,1]$, $\cos \phi \in [-1, 1]$, the potential must be separately minimised on the edges and vertices of that rectangle. The potential is required to be positive in these parameter regions in any case as a necessary condition.

The conditions for $\rho = 0$ (and any $\phi$) are given by \eqref{eq:ID:copos:crit:first}. The extremum solutions \eqref{eq:2HDM:noCPV:h1}, \eqref{eq:2HDM:noCPV:h2} and \eqref{eq:2HDM:noCPV:rho} are \emph{already} calculated for the edges $\cos \phi = \pm 1$ of the orbit space.  

On the edge $\rho = 1$, the other solution for $\phi$ becomes viable. The minimisation equations are \eqref{eq:2HDM:min:phi}, \eqref{eq:2HDM:min:h1}, \eqref{eq:2HDM:min:h2} and \eqref{eq:2HDM:min:lambda} with $\rho = 1$. The extremum solutions with $\cos \phi \neq \pm 1$ are given by
\begin{align}
  \cos \phi_{\rho = 1} &= -\frac{\lambda_{6} h_{1}^2 + \lambda_{7} h_{2}^2}
  {2 \lambda_{5} h_{1} h_{2}}, \label{eq:2HDM:noCPV:phi:rho:1}
  \\
  h_{1, \rho = 1}^{2} &= \frac{\lambda_{5} 
  (2 \lambda_{2} - \lambda_{3} - \lambda_{4} + \lambda_{5})  
  +  \lambda_{7} (\lambda_{6} - \lambda_{7})}
  {2 \lambda_{5} (\lambda_{1} + \lambda_{2} - \lambda_{3} - \lambda_{4} 
  + \lambda_{5}) - (\lambda_{6} - \lambda_{7})^2},
\label{eq:2HDM:noCPV:h1:rho:1}
  \\
  h_{2, \rho = 1}^{2} &= \frac{\lambda_{5} (2 \lambda_{1} - \lambda_{3} - \lambda_{4} 
  + \lambda_{5})  
  - \lambda_{6} (\lambda_{6} - \lambda_{7})}
  {2 \lambda_{5} (\lambda_{1} + \lambda_{2} - \lambda_{3} - \lambda_{4} 
  + \lambda_{5}) - (\lambda_{6} - \lambda_{7})^2},
\label{eq:2HDM:noCPV:h2:rho:1}
  \\
  V_{\text{min}, \rho = 1} &=  \frac{1}{2} \frac{
  4 \lambda_{1} \lambda_{2} \lambda_{5} -
  2 \lambda_{2} \lambda_{6}^2 - 2 \lambda_{1} \lambda_{7}^2
  - (\lambda_{3} + \lambda_{4} - \lambda_{5}) 
  [\lambda_{5} (\lambda_{3} + \lambda_{4} - \lambda_{5}) 
  - 2 \lambda_{6} \lambda_{7}] }
  {2 \lambda_{5} (\lambda_{1} + \lambda_{2} - \lambda_{3} - \lambda_{4} 
  + \lambda_{5}) - (\lambda_{6} - \lambda_{7})^2}.
\label{eq:2HDM:noCPV:V:rho:1}
\end{align}

In the vertices $\cos \phi = \pm 1, \rho = 1$ of the orbit space, the positivity conditions can be found by applying the conditions \eqref{eq:one:variable:polynomial:positivity} to 
$V_{\cos \phi = \pm 1, \rho = 1}/h_{2}^{4}$ with $x = h_{1}/h_{2}$. The discriminant $D$ and the polynomials $Q$ and $R$ are the same for $\phi = 0$ and for $\phi = \pi$ and given by
\begin{align}
  D_{\cos \phi = \pm 1,\rho=1} &= 16[ 16 \lambda_{1}^{3} \lambda_{2}^{3} + 
  \lambda_{1} \lambda_{2} \lambda_{345}^{4} - 
  27 \lambda_{2}^{2} \lambda_{6}^{4} - 
  48 \lambda_{1}^{2} \lambda_{2}^{2} \lambda_{6} \lambda_{7} - 
  6 \lambda_{1} \lambda_{2} \lambda_{6}^{2} \lambda_{7}^{2} - 
  16 \lambda_{6}^{3} \lambda_{7}^{3} \notag
  \\
  &- 27 \lambda_{1}^{2} \lambda_{7}^{4} - \lambda_{345}^{3} 
  (\lambda_{2} \lambda_{6}^{2} + \lambda_{1} \lambda_{7}^{2}) + 
  18 \lambda_{345} (2 \lambda_{1} \lambda_{2} + \lambda_{6} \lambda_{7}) 
  (\lambda_{2} \lambda_{6}^{2} + \lambda_{1} \lambda_{7}^{2}) 
  \\
  &+ \lambda_{345}^{2} (-8 \lambda_{1}^{2} \lambda_{2}^{2} 
  - 20 \lambda_{1} \lambda_{2} \lambda_{6} \lambda_{7} 
  + \lambda_{6}^{2} \lambda_{7}^{2})], \notag
  \\ 
  Q_{\cos \phi = \pm 1,\rho=1} &= 8 \lambda_{1} \lambda_{345} - 12 \lambda_{6}^{2},
  \\
  R_{\cos \phi = \pm 1,\rho=1} &= 16 [4 \lambda_{1}^{3} \lambda_{2} + 
   4 \lambda_{1} \lambda_{345} \lambda_{6}^{2} - 
   3 \lambda_{6}^{4} - \lambda_{1}^{2} (\lambda_{345}^{2} + 
      4 \lambda_{6} \lambda_{7})],
\end{align}
where $\lambda_{345} \equiv \lambda_{3} + \lambda_{4} + \lambda_{5}$.

Altogether, the conditions for the 2HDM potential with real couplings to be bounded from below are
\begin{equation}
\begin{split}
   &V_{\rho = 0}>0 \land D_{\cos \phi = \pm 1,\, \rho = 1} \land (Q_{\cos \phi = \pm 1,\, \rho = 1} > 0 \lor R_{\cos \phi = \pm 1,\, \rho = 1} > 0)
  \\
  &\land \left( 0 < h_{1, \rho = 1}^{2} < 1 \land 0 < h_{2, \rho = 1}^{2} < 1 
  \land 0 < \cos^{2} \phi_{\rho = 1} < 1 \implies V_{\text{min}, \rho = 1} > 0 \right)
  \\
  &\land \left( 0 < h_{1}^{2} < 1 \land 0 < h_{2}^{2} < 1 
  \land 0 < \rho^{2} < 1 \implies V_{\text{min}} > 0 \right),
\end{split}
\label{eq:2HDM:noCPV:vac:stab:cond}
\end{equation}
where the conditions for $V_{\rho = 0} > 0$ are given by \eqref{eq:ID:copos:crit:first} and $p \implies q$ is equivalent to $\lnot p \lor q$. In fact, it is enough to check that either $h_{1}^{2}$ or $h_{2}^{2}$ is within bounds, since they are related by $h_{1}^{2} + h_{2}^{2} = 1$. The conditions \eqref{eq:2HDM:noCPV:vac:stab:cond} are illustrated in Figure~\ref{fig:2HDM:parameter:space}.

The approach we use gives a simpler result than directly minimising $V_{\cos \phi = \pm 1, \rho = 1}$ with respect to $h_{1}$ and $h_{2}$ as we did for $V_\text{min}$, especially as the conditions \eqref{eq:one:variable:polynomial:positivity} automatically take into account the two different values $\cos \phi = \pm1$. For a potential that depends on three or more moduli of fields, e.g. the 3HDM, where we would have a $h_{3}$ besides $h_{1}$ and $h_{2}$, this is not possible and one has to minimise the potential on a hypersphere or use tensor eigenvalues (see Section \ref{sec:tensor:eigenvalues}).

\section{Vacuum Stability for $\mathbb{Z}_{3}$ Scalar Dark Matter}
\label{sec:z3:scalar:DM}

Another physical example is given by scalar dark matter stable under a $\mathbb{Z}_{3}$ discrete group. The most general scalar quartic potential of the SM Higgs $H_{1}$, an inert doublet $H_{2}$ and a complex singlet $S$ which is symmetric under a $\mathbb{Z}_{3}$ group is \cite{Belanger:2012vp,Belanger:2014bga}
\begin{equation}
\begin{split}
  V(H_{1}, H_{2}, S) &= \lambda_{1} \abs{H_{1}}^{4} + \lambda_{2} \abs{H_{2}}^{4} 
  + \lambda_{3} \abs{H_{1}}^{2} \abs{H_{2}}^{2} + \lambda_{4} (\hc{H_{1}} H_{2}) (\hc{H_{2}} H_{1})
  + \lambda_{S} \abs{S}^{4} \\
  &+ \lambda_{S1} \abs{S}^{2} \abs{H_{1}}^{2} 
  + \lambda_{S2} \abs{S}^{2} \abs{H_{2}}^{2}
  + \frac{1}{2} (\lambda_{S12} S^{2} \hc{H_{1}} H_{2} + \lambda_{S12}^{*} S^{\dagger 2} \hc{H_{2}} H_{1})
  \\
  &= \lambda_{1} h_{1}^{4} + \lambda_{2} h_{2}^{4} + \lambda_{3} h_{1}^{2} h_{2}^{2}
  + \lambda_{4} \rho^{2} h_{1}^{2} h_{2}^{2} + \lambda_{S} s^{4} + \lambda_{S1} s^{2} h_{1}^{2}
  + \lambda_{S2} s^{2} h_{2}^{2} - \abs{\lambda_{S12}} \rho s^{2} h_{1} h_{2}
  \\
  &\equiv \lambda_{S} s^{4} + M^{2}(h_{1}, h_{2}) s^{2} + V(h_{1}, h_{2}),
\end{split}
\label{eq:V:Z3:SIID}
\end{equation}
where we have used the parametrisation \eqref{eq:V:2HDM:param} for the doublet bilinears and $S = s e^{i \phi_{S}}$, and we have minimised $\cos (\phi + 2 \phi_{S} + \phi_{\lambda_{S12}}) = -1$ so $\lambda_{S12} = -\abs{\lambda_{S12}}$ without loss of generality. We define $M^{2}(h_{1}, h_{2}) \equiv \lambda_{S1} h_{1}^{2} - \abs{\lambda_{S12}} \rho h_{1} h_{2} + \lambda_{S2} h_{2}^{2}$ and $V(h_{1}, h_{2}) \equiv V(h_{1}, h_{2}, 0)$. 

The situation is similar to the case of two real scalars and the Higgs boson in Section~\ref{sec:two:real:scalars:and:Higgs}. First of all, $\lambda_{S} > 0$ and $V(h_{1}, h_{2}) > 0$. The conditions for $V(h_{1}, h_{2}) > 0$ are the same as in the inert doublet model with $\lambda_{5} = 0$:
\begin{gather}
  \lambda_{1} > 0,  \quad \lambda_{2} > 0, \quad \lambda_{3} + 2 \sqrt{\lambda_{1} \lambda_{2}} > 0, 
  \label{eq:ID:copos:crit:first:lambda:5:zero}
  \\
  \lambda_{3} + \lambda_{4} + 2 \sqrt{\lambda_{1} \lambda_{2}} > 0.
  \label{eq:ID:copos:crit:last:lambda:5:zero}
\end{gather}

We minimise the potential with respect to $h_{1}$, $h_{2}$, $s$ and $\rho$ with the fields lying on a sphere, enforced by a Lagrange multiplier $\lambda$. The minimisation equations are
\begin{align}
  h_{1} h_{2} \left( 2 \rho \lambda_{4} h_{1} h_{2} - \abs{\lambda_{S12}} s^{2} \right) &= 0,
  \\
  4 \lambda_{1} h_{1}^{3}  + 2 (\lambda_{3} + \lambda_{4} \rho^{2}) h_{1} h_{2}^{2} 
  + 2 \lambda_{S1} h_{1} s^{2} - \abs{\lambda_{S12}} \rho h_{2} s^{2}
  &= \lambda h_{1},
  \\
  4 \lambda_{2} h_{2}^{3}  + 2 (\lambda_{3} + \lambda_{4} \rho^{2}) h_{1}^{2} h_{2} 
  + 2 \lambda_{S2} h_{2} s^{2} - \abs{\lambda_{S12}} \rho h_{1} s^{2}
  &= \lambda h_{2},
  \\
  s \left( 4 \lambda_{S} s^{2} + 2 \lambda_{S1} h_{1}^{2} + 2 \lambda_{S2} h_{2}^{2} 
  - 2 \abs{\lambda_{S12}} \rho h_{1} h_{2} \right) &= \lambda s,
  \\
  h_{1}^{2} + h_{2}^{2} + s^{2} &= 1.
\end{align}
The solution with all fields non-zero is
\begin{align}
  \rho &= \Big(\abs{\lambda_{S12}} s^{2} \Big) \Big/ \Big( 2 \lambda_{4} h_{1} h_{2} \Big),
\label{eq:Z3:SIID:rho}
  \\
  \label{eq:Z3:SIID:h1}
  h_{1}^{2} &= \frac{1}{2} \Big( (2 \lambda_{2} - \lambda_{3}) 
  (4 \lambda_{S} \lambda_{4} - \abs{\lambda_{S12}}^{2}) 
  + 2 \lambda_{4} [(\lambda_{3} + \lambda_{S1}) \lambda_{S2} - 2 \lambda_{2} \lambda_{S1} 
  - \lambda_{S2}^{2}] \Big)  \Big/ \Big( (\lambda_{1} + \lambda_{2} - \lambda_{3})
  \\
  &\times (4 \lambda_{S} \lambda_{4} - \abs{\lambda_{S12}}^{2})
  + \lambda_{4} [4 \lambda_{1} \lambda_{2} - \lambda_{3}^{2} 
  - 4 \lambda_{1} \lambda_{S2} - 4 \lambda_{2} \lambda_{S1} + 2 \lambda_{3} (\lambda_{S1} + \lambda_{S2}) - (\lambda_{S1} - \lambda_{S2})^{2}] \Big),
  \notag
  \\
\label{eq:Z3:SIID:h2}
  h_{2}^{2} &= \frac{1}{2} \Big( (2 \lambda_{1} - \lambda_{3}) 
  (4 \lambda_{S} \lambda_{4} - \abs{\lambda_{S12}}^{2}) 
  + 2 \lambda_{4} [(\lambda_{3} + \lambda_{S2}) \lambda_{S1} - 2 \lambda_{1} \lambda_{S2} 
  - \lambda_{S1}^{2}] \Big)  \Big/ \Big( (\lambda_{1} + \lambda_{2} - \lambda_{3})
  \\
  &\times (4 \lambda_{S} \lambda_{4} - \abs{\lambda_{S12}}^{2})
  + \lambda_{4} [4 \lambda_{1} \lambda_{2} - \lambda_{3}^{2} 
  - 4 \lambda_{1} \lambda_{S2} - 4 \lambda_{2} \lambda_{S1} + 2 \lambda_{3} (\lambda_{S1} + \lambda_{S2}) 
  - (\lambda_{S1} - \lambda_{S2})^{2}] \Big),
  \notag
  \\
  \label{eq:Z3:SIID:s}
  s^{2} &= \lambda_{4} \Big( 4 \lambda_{1} \lambda_{2} - \lambda_{3}^{2} - 2 \lambda_{1} \lambda_{S2} - 2 \lambda_{2} \lambda_{S1} + \lambda_{3} (\lambda_{S1} + \lambda_{S2}) \Big)  \Big/ \Big( (\lambda_{1} + \lambda_{2} - \lambda_{3})
  \\
  &\times (4 \lambda_{S} \lambda_{4} - \abs{\lambda_{S12}}^{2})
  + \lambda_{4} [4 \lambda_{1} \lambda_{2} - \lambda_{3}^{2} 
  - 4 \lambda_{1} \lambda_{S2} - 4 \lambda_{2} \lambda_{S1} 
  + 2 \lambda_{3} (\lambda_{S1} + \lambda_{S2}) - (\lambda_{S1} - \lambda_{S2})^{2}] \Big),
  \notag
\\
\label{eq:Z3:SIID:V:min}
  V_{\rm min} &= \frac{1}{4} \Big( (4 \lambda_{1} \lambda_{2} 
  - \lambda_{3}^{2}) (4 \lambda_{S} \lambda_{4} - \abs{\lambda_{S12}}^{2}) 
  - 4 \lambda_{4} (\lambda_{1} \lambda_{S2}^{2} + \lambda_{2} \lambda_{S1}^{2} - \lambda_{3} \lambda_{S1} \lambda_{S2}) \Big)  \Big/ 
  \Big( (\lambda_{1} + \lambda_{2} - \lambda_{3})
  \\
  &\times (4 \lambda_{S} \lambda_{4} - \abs{\lambda_{S12}}^{2})
  + \lambda_{4} [4 \lambda_{1} \lambda_{2} - \lambda_{3}^{2} 
  - 4 \lambda_{1} \lambda_{S2} - 4 \lambda_{2} \lambda_{S1} + 2 \lambda_{3} (\lambda_{S1} +  \lambda_{S2}) - (\lambda_{S1} - \lambda_{S2})^{2}] \Big).
  \notag
\end{align}
Note that $h_{1}^{2}$, $h_{2}^{2}$, $s^{2}$ and $V_{\rm min}$ share the same denominator. The solution $s = 0$ will repeat \eqref{eq:ID:copos:crit:first:lambda:5:zero} and \eqref{eq:ID:copos:crit:last:lambda:5:zero} and the solutions $h_{1} = 0$ or $h_{2} = 0$ will be made redundant by eq. \eqref{V:Z3:SIID:crit:rho:0}.

Since $\rho = 0$ sets the $\lambda_{S12}$ term to zero, $V_{\rho = 0} > 0$ is biquadratic in the fields and we can calculate the positivity conditions for $V_{\rho = 0} > 0$ via (strict) copositivity of the matrix of couplings in the $(h_{1}^{2}, h_{2}^{2}, s^{2})$ basis,
\begin{align}
  \lambda_{S} &> 0, & \lambda_{1} &> 0, & \lambda_{2} &> 0,
  \notag
  \\
  \bar{\lambda}_{3} \equiv \lambda_{3} + 2 \sqrt{\lambda_{1} \lambda_{2}} &> 0,
  &
  \bar{\lambda}_{S1} \equiv \lambda_{S1} + 2 \sqrt{\lambda_{S} \lambda_{1}} &> 0,
  &
  \bar{\lambda}_{S2} \equiv \lambda_{S2} + 2 \sqrt{\lambda_{S} \lambda_{2}} &> 0,
  \label{V:Z3:SIID:crit:rho:0}
  \\
  \span \span \span \span \sqrt{\lambda_{S}} \lambda_{3} + \sqrt{\lambda_{1}} \lambda_{S2} + \sqrt{\lambda_{2}} \lambda_{S1} 
  + \sqrt{\lambda_{S} \lambda_{1} \lambda_{2}} 
  + \sqrt{\bar{\lambda}_{S1} \bar{\lambda}_{S2} \bar{\lambda}_{3}} &> 0,
  \notag
\end{align}
partly repeating conditions \eqref{eq:ID:copos:crit:first:lambda:5:zero}.

For $\rho = 1$, instead of direct minimisation of the potential in all variables, we find it easier to calculate conditions very similar to the case \eqref{eq:V:general:2:fields:H:vac:stab} of two real scalars and the Higgs boson. To reduce $V_{\rho = 1}$ to a polynomial of two variables, we minimise it with respect to $s^{2}$:
\begin{equation}
  s^{2}_{\text{min}} = -\frac{1}{2 \lambda_{S}} M^{2}(h_{1}, h_{2}).
\end{equation}

Again, either $M^{2}(h_{1}, h_{2}) > 0$ and the solution for $s^{2}_{\rm min}$ is unphysical, or else $V_{s^{2} = s^{2}_{\rm min}} = V(h_{1}, h_{2}) - \frac{1}{4 \lambda_{S}} M^{4}(h_{1}, h_{2}) > 0$. We have
\begin{equation}
\begin{split}
  4 \lambda_{S} V_{s^{2} = s^{2}_{\rm min}} &= (4 \lambda_{1} \lambda_{S} - \lambda_{S1}^{2}) h_{1}^{4}
  + (4 \lambda_{2} \lambda_{S} - \lambda_{S2}^{2}) h_{2}^{4} 
  + \left[ 4(\lambda_{3} 
  + \lambda_{4}) \lambda_{S} - 2 \lambda_{S1} \lambda_{S2} - \abs{\lambda_{S12}}^{2} \right] h_{1}^{2} h_{2}^{2} 
  \\
  &+ \lambda_{S1} \abs{\lambda_{S12}} h_{1}^{3} h_{2} + \lambda_{S2} \abs{\lambda_{S12}} h_{1} h_{2}^{3}.
\end{split}
\label{eq:V:Z3:SIID:s:min}
\end{equation}
The coefficient matrix of $M^{2}$ is given by
\begin{equation}
  \mathbf{M}^{2} = 
  \begin{pmatrix}
    \lambda_{S1} & -\frac{1}{2} \abs{\lambda_{S12}} \\
    -\frac{1}{2} \abs{\lambda_{S12}} & \lambda_{S2}
  \end{pmatrix}.
\end{equation}

\begin{figure}[tb]
\begin{center}
\includegraphics{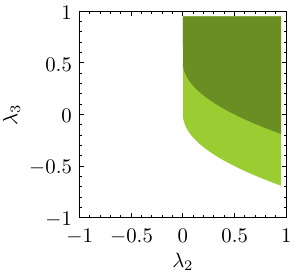}
\includegraphics{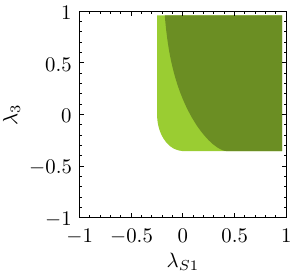}
\includegraphics{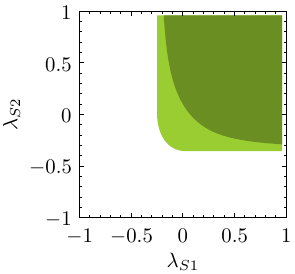}
\\
\vspace{4mm}
\includegraphics{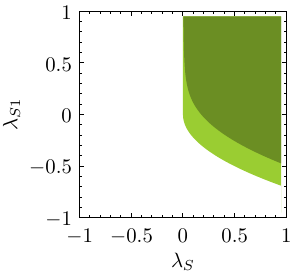}
\includegraphics{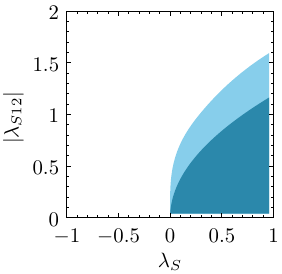}
\includegraphics{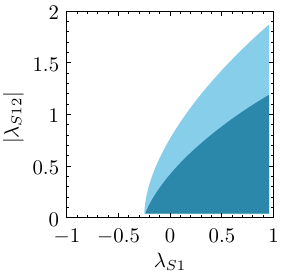}
\caption{An example of parameter space allowed by the vacuum stability conditions \eqref{eq:Z3:SIID:conditions} for the potential \eqref{eq:V:Z3:SIID} of the SM Higgs, an inert doublet and a complex singlet. The values of parameters are $\lambda_{H} = 0.125$, $\lambda_{S} = 0.125$, $\lambda_{2} = 0.25$, $\lambda_{3} = \lambda_{4} = \lambda_{S1} = \lambda_{S2} = 0$, except for plots where these couplings vary. Both light and dark green regions are allowed for for $\abs{\lambda_{S12}} = 0$, while $\abs{\lambda_{S12}} = 0.5$ only allows the dark green region. In the last two panels, where $\abs{\lambda_{S12}}$ varies, only the dark blue region is allowed for $\lambda_{S2} = 0$, while $\lambda_{S2} = 0.5$ allows both the dark and light blue regions.}
\label{fig:12S:plots}
\end{center}
\end{figure}

Repeating the procedure of Section \ref{sec:two:real:scalars:and:Higgs} step by step, the conditions for $V_{\rho = 1} > 0$, in addition to $\lambda_{S} > 0$ and \eqref{eq:ID:copos:crit:first:lambda:5:zero} and \eqref{eq:ID:copos:crit:last:lambda:5:zero}, are given by 
\begin{equation}
\begin{split}
  & \lambda_{S1} \leqslant 0 \land \lambda_{S2} \leqslant 0 
  \land \abs{\lambda_{S12}}^{2} \leqslant 4 \lambda_{S1} \lambda_{S2}
  \implies 
  4 \lambda_{S} \lambda_{1} - \lambda_{S1}^{2} > 0 
  \land 4 \lambda_{S} \lambda_{2} - \lambda_{S2}^{2} > 0 
  \\
  & \land D_{s^{2} = s^{2}_{\rm min}} \land 
  (Q_{s^{2} = s^{2}_{\rm min}} > 0 \lor R_{{s}^{2} = s^{2}_{\rm min}}
   > 0),
  \\
  & \abs{\lambda_{S12}}^{2} > 4 \lambda_{S1} \lambda_{S2} 
  \implies
  \lambda'_{40} > 0 \land \lambda'_{04} > 0
  \\
  &\land \left[ \left( D_{s^{2} = s^{2}_{\rm min}} \leqslant 0 \land 
  \left( \lambda'_{31} \sqrt{\lambda'_{04}} + \lambda'_{13} \sqrt{\lambda'_{40}} \right) > 0 \right) \right. \\
  &\lor 
  \left( -2 \sqrt{\lambda'_{40} \lambda'_{04}} < \lambda'_{22} < 6 \sqrt{\lambda'_{40} \lambda'_{04}} \land D_{s^{2} = s^{2}_{\rm min}} \geqslant 0 
  \land \Lambda'_{1 \: s^{2} = s^{2}_{\rm min}} \leqslant 0 \right) 
  \\
  &\lor 
  \left( 6 \sqrt{\lambda'_{04} \lambda'_{40}} < \lambda'_{22} \land 
  [(\lambda'_{13} > 0 \land \lambda'_{31} > 0) 
 \left. \lor (D_{s^{2} = s^{2}_{\rm min}} \geqslant 0 
  \land \Lambda'_{2  \: s^{2} = s^{2}_{\rm min}} \leqslant 0) ] \right) \right],
\end{split}
\label{V:Z3:SIID:crit:rho:1}
\end{equation}
where we  have taken $\lambda'_{31} \to -\sgn (\lambda_{S2} - \lambda_{S1}) \lambda'_{31}$ and $\lambda'_{13} \to -\sgn (\lambda_{S2} - \lambda_{S1}) \lambda'_{13}$. Similarly to Section \ref{sec:two:real:scalars:and:Higgs}, we denote by $\lambda'_{ij}$  the coefficients of the transformed potential $V'_{s^{2} = s^{2}_{\rm min}}$, in the basis where $M^{\prime 2}$ is anti-diagonal.\footnote{In eq. \eqref{V:Z3:SIID:crit:rho:1}, we have retained the $\lambda'_{ij}$ notation of Section \ref{sec:two:real:scalars:and:Higgs} for ease of comparison. In the 2HDM notation, the coefficients would be $\lambda'_{40} \equiv \lambda'_{1}$, $\lambda'_{04} \equiv \lambda'_{2}$, $\lambda'_{22} \equiv \lambda'_{345}$, $\lambda'_{31} \equiv \lambda'_{6}$ and $\lambda'_{13} \equiv \lambda'_{7}$.} Note that we are justified to use the conditions \eqref{eq:one:variable:polynomial:positivity} for the positivity of the general quartic on reals in the second line of \eqref{V:Z3:SIID:crit:rho:1}, because in \eqref{eq:V:Z3:SIID:s:min} the coefficients of the $h_{1}^{3} h_{2}$ and $h_{1} h_{2}^{3}$ terms are non-positive in that case.

Altogether, the conditions for the potential \eqref{eq:V:Z3:SIID} symmetric under a $\mathbb{Z}_{3}$  to be bounded from below are
\begin{equation}
\begin{split}
  &\lambda_{S} > 0 \land V(h_{1}, h_{2}) > 0 \land V_{\rho = 0} > 0 \land V_{\rho = 1} > 0 
  \\
  &\land \left( 0 < h_{1}^{2} < 1 \land 0 < h_{2}^{2} < 1 
  \land 0 < s^{2} < 1 
  \land 0 < \rho^{2} < 1 \implies V_{\text{min}} > 0 \right),
\end{split}
\label{eq:Z3:SIID:conditions}
\end{equation}
where the conditions for $V(h_{1}, h_{2}) > 0$ are given by \eqref{eq:ID:copos:crit:first:lambda:5:zero} and \eqref{eq:ID:copos:crit:last:lambda:5:zero}, the conditions for $V_{\rho = 0} > 0$ are given by \eqref{V:Z3:SIID:crit:rho:0}, the conditions for $V_{\rho = 1} > 0$ are given by \eqref{V:Z3:SIID:crit:rho:1}, and the extremum solutions $h_{1}^{2}$, $h_{2}^{2}$, $s^{2}$, $\rho^{2}$ and $V_{\rm min}$ are given by \eqref{eq:Z3:SIID:rho}, \eqref{eq:Z3:SIID:h1}, \eqref{eq:Z3:SIID:h2}, \eqref{eq:Z3:SIID:s} and \eqref{eq:Z3:SIID:V:min}. The conditions \eqref{eq:Z3:SIID:conditions} are illustrated in Figure~\ref{fig:12S:plots}.

\section{Tensor Eigenvalues}
\label{sec:tensor:eigenvalues}

\subsection{Positive Tensors}

The most general scalar potential of $n$ real singlet scalar fields $\phi_{i}$ can be written as
\begin{equation}
  V = \lambda_{ijkl} \phi_i \phi_j \phi_k \phi_l \equiv \Lambda \phi^{4},
\end{equation}
where $\Lambda$ is the tensor of scalar couplings and $\phi = (\phi_{1}, \ldots, \phi_{n})$. Clearly, the tensor $\Lambda$ can always made fully symmetric under permutations of the indices of its elements.

A symmetric matrix is positive if its eigenvalues are greater than zero. Can one generalise matrix eigenvalues and eigenvectors to tensors in such a way that they have similar properties? Indeed, tensor eigenvalues and eigenvectors have been defined independently by Qi in \cite{Qi20051302} (which we follow in our exposition) and Lim \cite{2006math......7648L}.

A real $m$th-order $n$-dimensional tensor $A$ has $n^{m}$ elements $A_{i_{1} \ldots i_{m}}$, where $i_{j} = 1, \ldots, n$ for $j = 1, \ldots, m$. A homogenous polynomial $f(x)$ of $n$ variables and degree $m$ can be written as the tensor product
\begin{equation}
  f(x) = A x^{m} \equiv A_{i_{1} \ldots i_{m}} x_{i_{1}} \cdots x_{i_{m}}.
\end{equation}
In analogy with non-negative matrices, \emph{an $n$-dimensional tensor $A$ is called non-negative if $A x^{m} \geqslant 0$ for all $x \in \mathbb{R}^{n}$.}

For a vector $x \in \mathbb{R}^{n}$ we define $(x^{[m]})_{i} = x_{i}^{m}$. The number $\lambda$ is an eigenvalue of $A$ if it is a solution to the equation
\begin{equation}
  A x^{m-1} = \lambda x^{[m-1]},
  \label{eq:tensor:eigenvalue:eq}
\end{equation}
where $x$ is an eigenvector of $A$. For $m = 2$, the tensor $A$ is a matrix and the equation \eqref{eq:tensor:eigenvalue:eq} coincides with the usual matrix eigenvalue equation. When $m > 2$, the  eigenvalues and eigenvectors of a tensor can be complex, but for even $m$, there always exist real eigenvalues and eigenvectors, so-called $H$-eigenvalues and $H$-eigenvectors.

An $n$-dimensional symmetric tensor of order $m$ has $n (m-1)^{n-1}$ eigenvalues, so a coupling tensor of a renormalisable potential of $n$ real scalar fields in 4-dimensional spacetime has $n \, 3^{n-1}$ eigenvalues: for example, with two real fields, there are six eigenvalues. The product of all the eigenvalues of $A$ is the hyperdeterminant $\det A$, that is the resultant of $A x^{m-1} = 0$. When $m = 2$, the hyperdeterminant reduces to the usual matrix determinant. The sum of all eigenvalues is $(m - 1)^{n-1}$ times the sum of diagonal elements or trace $\tr{A} = \sum_{i} A_{iiii}$.

For even $m$, an eigenvalue equation similar to the matrix eigenvalue equation can be given,
\begin{equation}
  \det (A - \lambda I) = 0,
\label{eq:tensor:H:eigenvalues}
\end{equation}
where $I$ is the unit tensor with $I_{i_{1}\ldots i_{m}} = \delta_{i_{1}\ldots i_{m}}$. The degree of the eigenvalue equation is $d = n (m - 1)^{n-1}$.

Obviously only real eigenvectors have relevance to whether the tensor $A$ is non-negative. \emph{The tensor $A$ is non-negative if all its eigenvalues with real eigenvectors are non-negative.} All the principal subtensors of $A$, that is the tensors obtained by setting one or more of the variables $x_{i}$ in $f(x)$ to zero, must be positive as well.

Qi also defines tensor $E$-eigenvalues and $E$-eigenvectors, which are exactly the same as solutions to the equations \eqref{eq:eqs:Lagrange:multiplier} with fields constrained to a hypersphere. For even $m$, there always exist real $E$-eigenvalues and $E$-eigenvectors.

Copositive tensors are defined in obvious analogy to copositive matrices \cite{QI2013228}: \emph{A real symmetric tensor $A$ of order $m$ and dimension $n$ is copositive if $A x^{m} \geqslant 0 \text{ for all } x \in \mathbb{R}_{+}^{n}$.} It has been shown that Kaplan's test of matrix co-positivity \cite{Kaplan:2000:TCM} directly generalises to copositive tensors \cite{2013arXiv1302.6084S}: \emph{A symmetric tensor $A$ is copositive if and only if every principal subtensor of $A$ has no eigenvector $v > 0$ with associated H-eigenvalue $\lambda < 0$.} 

\subsection{Vacuum Stability of the Potential of Two Real Scalars}

As an example, we consider again the general potential of two real scalars given by \eqref{eq:V:general:2:fields}:
\begin{equation}
  V(\phi_{1}, \phi_{2}) = \lambda_{ij} \phi_{1}^{i} \phi_{2}^{j} 
  = \lambda_{40} \phi_{1}^{4} + \lambda_{31} \phi_{1}^{3} \phi_{2} + \lambda_{22} \phi_{1}^{2} \phi_{2}^{2} + \lambda_{13} \phi_{1} \phi_{2}^{3} + \lambda_{04} \phi_{2}^{4}.
\end{equation}

The tensor of the scalar couplings of the potential is given by
\begin{equation}
  \Lambda = 
  \begin{pmatrix}
    \begin{pmatrix}
      \lambda_{40} & \frac{1}{4} \lambda_{31} \\
      \frac{1}{4} \lambda_{31} & \frac{1}{6} \lambda_{22}
    \end{pmatrix}
    &
    \begin{pmatrix}
      \frac{1}{4} \lambda_{31} & \frac{1}{6} \lambda_{22} \\
      \frac{1}{6} \lambda_{22} & \frac{1}{4} \lambda_{13}
    \end{pmatrix}
    \\
     \begin{pmatrix}
      \frac{1}{4} \lambda_{31} & \frac{1}{6} \lambda_{22} \\
      \frac{1}{6} \lambda_{22} & \frac{1}{4} \lambda_{13}
    \end{pmatrix}
    &
    \begin{pmatrix}
      \frac{1}{6} \lambda_{22} & \frac{1}{4} \lambda_{13} \\
      \frac{1}{4} \lambda_{13} & \lambda_{04}
    \end{pmatrix}
  \end{pmatrix},
\end{equation}
that is, $\lambda_{1111} = \lambda_{40}$, $\lambda_{2222} = \lambda_{04}$, $\lambda_{1112} = \lambda_{1121} = \lambda_{1211} = \lambda_{2111} = \frac{1}{4} \lambda_{31}$ and so on.

The $H$-eigenvalue equations are
\begin{equation}
\begin{split}
  4 \lambda_{40} \phi_{1}^{3} + 3 \lambda_{31} \phi_{1}^{2} \phi_{2} + 
  2 \lambda_{22} \phi_{1} \phi_{2}^{2} + \lambda_{13} \phi_{2}^{3} &= 4 \lambda \phi_{1}^{3},
  \\
  \lambda_{31} \phi_{1}^{3} + 2 \lambda_{22} \phi_{1}^{2} \phi_{2} + 
  3 \lambda_{13} \phi_{1} \phi_{2}^{2} + 4 \lambda_{04} \phi_{2}^{3} &= 4 \lambda \phi_{2}^{3},
 \end{split}
\label{eq:tensor:eval:eqs}
\end{equation}

The product of all solutions to the $H$-eigenvalue equations \eqref{eq:tensor:eval:eqs} is exactly the discriminant $D$ given by \eqref{2:real:2:field:discriminant} and their sum is $3 (\lambda_{40} + \lambda_{04})$. 
The tensor eigenvalue equation \eqref{eq:tensor:H:eigenvalues} is a 6th degree equation, and the eigenvalues cannot in general be solved in radicals. Therefore, for two fields, we are better off using the conditions \eqref{eq:2:real:2:field:vac:stab}. 

But for the general potential of three fields we have no such (relatively) simple conditions. For $m = 4$, $n = 3$, the recipe for the hyperdeterminant, given in \cite{4608938}, is very complicated in practice. Analytical expressions can be found only for coupling tensors of potentials that have few fields and are rather symmetric.   The equations \eqref{eq:tensor:eigenvalue:eq}, however, can be easily solved numerically (taking e.g. $x_{1} = 1$ without loss of generality).

\section{Conclusions}
\label{sec:conclusions}

In particle physics, scalar potentials have to be bounded from below in order for the physics to make sense. Finding such conditions is a hard problem of algebraic geometry. We present analytical necessary and sufficient vacuum stability conditions for potentials of a few fields, where `few' means two or more, depending on field content and symmetry. The vacuum stability conditions \eqref{eq:2:real:2:field:vac:stab} for a general potential of two real fields fit on a few lines. Already for three fields, practical analytical conditions \eqref{eq:V:general:2:fields:H:vac:stab} can only be found for the potential \eqref{eq:V:general:2:fields:H}, where at least one of the fields, such as the Higgs boson, is present only in biquadratic form. In this case the problem reduces to the positivity of a general quartic polynomial with a quadratic constraint.

As further examples that put several of the discussed techniques to use, we present simple vacuum stability conditions \eqref{eq:2HDM:noCPV:vac:stab:cond} for the 2HDM potential without explicit CP-violation, and vacuum stability conditions \eqref{eq:Z3:SIID:conditions} for $\mathbb{Z}_{3}$ scalar dark matter with an inert doublet and a complex singlet. All analytical calculations have been checked numerically. We include a Mathematica notebook with the conditions with the \LaTeX\ source of the paper.

The vacuum stability conditions for the general potential of two real singlets (without or with the Higgs boson), and for the $\mathbb{Z}_{3}$ scalar dark matter are novel results. The vacuum stability conditions for the 2HDM potential with real couplings are in a shorter form than previous similar results \cite{Eriksson:2009ws}.

Of course, our endeavour can be made much easier if a restrictive symmetry is imposed on the potential. If all the fields  appear solely quadratically, for example, the problem becomes much simpler. Then the bounded from below conditions are given by copositivity constraints of the matrix of couplings. Many potentials can be written in terms of magnitudes of squares of fields and a few orbit space parameters.

The parameter space for more complicated potentials must be found numerically by minimising the potential on a hypersphere of field values or solving tensor eigenvalue equations. Still, some insight can be gained from (necessary) analytical conditions for a subspace where a field or more is set to zero, and the introduced methods can be used to reduce the parameter space for a numerical scan.

\subsection*{Acknowledgements}

We would like to thank Hardi Veerm\"{a}e for useful discussions. This work was supported by grants the Estonian Research Council grant PUT799, the grant IUT23-6 of the Estonian Ministry of Education and Research, and by the EU through the ERDF CoE program.

\bibliographystyle{jhep}
\bibliography{general}
\end{document}